%% file: DREAMpaper.tex
\documentclass[preprint,3p,twocolumn]{elsarticle}

\usepackage{amsmath}
\usepackage{amssymb}
\usepackage{url}
\usepackage{color}
\usepackage{siunitx}
\usepackage{booktabs}

\input{defs.tex}

\newcounter{bla}

\journal{Computer Physics Communications}

\begin{document}

\begin{frontmatter}
    \title{DREAM: a fluid-kinetic framework for tokamak disruption runaway electron simulations}

    \author{Mathias Hoppe\corref{author}}
    \author{Ola Embreus}
    \author{T\"unde F\"ul\"op}

    \cortext[author]{\textit{E-mail address:} hoppe@chalmers.se}
    \address{Department of Physics, Chalmers University of Technology, Gothenburg, SE-41296, Sweden}

    \begin{abstract}
Avoidance of the harmful effects of runaway electrons (REs) in
plasma-terminating disruptions is pivotal in the design of safety
systems for magnetic fusion devices.  Here,
we describe a computationally efficient numerical tool, that allows for
self-consistent simulations of plasma cooling and associated RE
dynamics during disruptions. It solves flux-surface averaged transport
equations for the plasma density, temperature and poloidal flux, using
a bounce-averaged kinetic equation to self-consistently provide the
electron current, heat, density and RE evolution, as well as the
electron distribution function. As an example, we consider disruption
scenarios with material injection and compare the electron dynamics
resolved with different levels of complexity, from fully kinetic to
fluid modes.

    \end{abstract}

    \begin{keyword}
    \end{keyword}
\end{frontmatter}

{\bf PROGRAM SUMMARY}\\
\begin{small}
    \noindent
    {\em Program Title:} \DREAM                                   \\
    {\em CPC Library link to program files:} (to be added by Technical Editor) \\
    {\em Developer's repository link:} \url{https://github.com/chalmersplasmatheory/DREAM} \\
    {\em Code Ocean capsule:} (to be added by Technical Editor)\\
    {\em Licensing provisions:} MIT\\
        {\em Programming language:} C++, Python                       \\
    {\em Nature of problem:}     Self-consistently simulates the plasma evolution in a tokamak disruption, with
    specific emphasis on runaway electron dynamics. The runaway
    electrons can be simulated either as a fluid, fully kinetically, or as a
    mix of the two. Plasma temperature, current density, electric field, ion density and
     charge states are all evolved self-consistently, where kinetic non-thermal contributions
    are captured using an orbit-averaged relativistic electron Fokker-Planck equation, which 
    couples to the plasma evolution. In the typical use case,
    the electrons are represented by two distinct populations: a cold fluid population and a kinetic superthermal population.\\
    {\em Solution method:}     The system of equations is solved using a standard multidimensional Newton's
    method. Partial differential equations---most prominently the bounce-averaged
    Fokker--Planck and current diffusion equations---are discretized using a
    high-resolution finite volume scheme that preserves density and
    positivity.
 \end{small}

\section{Introduction}
Disruptions of tokamak plasmas involve a partial loss of magnetic
confinement and a sudden cooling of the plasma \cite{Hender2007}. This
thermal quench leads to an increase in the plasma resistivity, causing
the plasma current to decay over a period termed the current
quench. The toroidal current cannot change significantly on the short
thermal quench timescale, and therefore an inductive electric field is
produced that can lead to electron runaway.  The main runaway
generation processes are the Dreicer \cite{Dreicer1959,Connor1975},
hot-tail \cite{Chiu1998,Smith2005,Aleynikov2017} and avalanche
\cite{sokolov1979multiplication,RosenbluthPutvinski1997,olaknockon2018}
mechanisms. In the nuclear phase of operations, these are complemented
with electrons generated by the beta decay of tritium and Compton scattering
of $\gamma$-rays emitted by the activated wall.

During the current quench, a large part of the plasma
current can be converted to a beam of energetic electrons which has
the potential to cause severe damage to plasma facing components
\cite{lehnen15disruptions,Breizman_2019}.  As the runaway generation
is exponentially sensitive to the plasma current
\cite{RosenbluthPutvinski1997}, this problem is expected to be
unacceptable in future high current tokamaks, such as ITER
\cite{Hender2007} and SPARC \cite{Sweeney2020}. Development of control
methods for avoidance or mitigation of disruptions is therefore of
critical importance and urgency in fusion physics.

The most discussed disruption mitigation method is material
injection. However, in certain cases the injection of impurities and
the associated radiative cooling can lead to the generation of even
larger runaway currents
 \cite{Vallhagen2020}.  There are a large number of degrees of
freedom associated with proposed mitigation methods
\cite{HollmannDMS}. However, only a small part of the parameter space
is accessible to existing experiments and the extrapolation of the results of existing experiments to next-generation tokamaks is not straightforward \cite{Boozer_2017,boozer2018pivotal}. Therefore theoretical modelling
and computationally efficient numerical simulations of the disruption
and associated runaway electron generation are essential.

Nonlinear magnetohydrodynamic (MHD) simulation tools, such as {\sc Jorek}
\cite{jorek1,jorek2} and {\sc Nimrod} \cite{nimrod} have many of the necessary
components to simulate mitigated disruptions. However, they so far
only allow studies of the dynamics of runaway electrons as test
particles \cite{Sommariva_2017,Sommariva_2018} or via a fluid model
\cite{Cai_2015,Matsuyama_2017,Bandaru_2019}, and are computationally expensive.
To self-consistently simulate disruption dynamics, an integrated tool that can
simulate situations when relativistic electrons comprise a significant part of
the electron distribution is required.

Simplified fluid codes such as {\sc go} \cite{Vallhagen2020}
and the 1.5-dimensional transport toolkit {\sc astra-strahl}
\cite{Linder_2020} include kinetically benchmarked models for Dreicer
and avalanche runaway electron generation, but have simplified models
for hot-tail generation \cite{Smith2008}.  Fully kinetic tools modelling the
momentum-space dynamics of relativistic electrons, such as {\sc code}
\cite{Landreman2014,Stahl2016}, {\sc norse} \cite{norse}, or the
bounce-averaged codes {\sc luke} \cite{Decker2004} and {\sc cql3d}
\cite{Chiu1998,Harvey2000} are suitable for capturing the hot-tail generation,
however, to couple these codes to a self-consistent model of the global
disruption dynamics leads to prohibitively expensive simulations.

In this paper we describe a new integrated tool for self-consistently
simulating the evolution of temperature, poloidal flux, and impurity
densities, along with the generation and transport of runaway electrons in
tokamaks: \DREAM\ (Disruption Runaway Electron Analysis Model). The
fully-implicit tool solves a nonlinear set of coupled equations
describing the evolution of temperature, density, current density and
electric field, as well as the full electron distribution function in
arbitrary axisymmetric geometry. It employs a combination of fluid
models for background plasma parameters, including the toroidal
electric field, electron and ion temperatures, ion densities and
charge states, as well as various models for runaway electrons, ranging from
fluid to fully kinetic. The most complete drift-kinetic model includes a
fully relativistic Fokker-Planck test-particle operator for
electron-electron collisions, 
synchrotron radiation reaction, an avalanche operator, bremsstrahlung and effects of screening in a
partially ionized plasma. Neglecting the field-particle part of the collision operator means that the conductivity is underestimated \cite{helander},   therefore  the  ohmic current is amended with a 
conductivity correction to capture the correct Spitzer response to an electric field.
A distinguishing feature of \DREAM\ is the
possibility of choosing reduced kinetic modes, which allow parts of
the electron phase space to be modelled kinetically, and the remainder
to be described by fluid equations.

The equations that describe the electron kinetics and background
plasma evolution in \DREAM\ are discussed in
Sections~\ref{sec:theory}-\ref{sec:plasma transport}. The numerical implementation is
then outlined in Section~\ref{sec:implementation}, and benchmarked
in Section~\ref{sec:tests} through a comparison to other numerical tools in
various limits. Finally, in Section~\ref{sec:examples}, we use \DREAM\ to
investigate a disruption simulation in a toroidal plasma, through which we
highlight the difference between the hierarchy of electron models implemented.

\section{Electron kinetics}\label{sec:theory}
\input{theory.tex}

\section{Background plasma evolution}\label{sec:plasma transport}
\input{backgroundplasma.tex}

\section{Numerics and implementation}\label{sec:implementation}
\input{implementation.tex}

\section{Tests \& benchmarks}\label{sec:tests}
\input{tests.tex}

\section{Comparison of electron models}\label{sec:examples}
\input{examples.tex}

\section{Summary}
The main purpose of the \DREAM\ code is to model the
self-consistent plasma evolution and runaway electron generation during
a tokamak disruption. The output is the evolution of the
temperature, densities of the different particle species and poloidal
flux (which sets the evolution of the current density) as well
as the electron distribution function.
The temperature evolution includes ohmic heating, radiated
power using atomic rate coefficients, collisional energy transfer from
hot electrons and ions, as well as dilution cooling.  The poloidal flux
evolution includes the option to model rapid current flattening
associated with fast magnetic reconnection events, via a
helicity-conserving hyperresistivity term. The fluid quantities are
solved on a one-dimensional flux-surface averaged grid, and the
kinetic equation for the electron distribution is solved in a three-dimensional
(1D-2P) bounce-averaged formulation.

The ability to treat electrons at various degrees of sophistication is one of
the more novel contributions of \DREAM. The physics of tokamak disruptions
typically involves multiple temporal and spatial scales, and so far
often required comprehensive and computationally expensive simulations involving
both fluid and full kinetic physics. By separating the electrons into cold, hot
and runaway populations, and evolving the cold and runaway electrons using fluid
models, \DREAM\ avoids resolving the usually uninteresting---but computationally
intensive---kinetic bulk and runaway tail dynamics. Furthermore, the possibility
to evolve hot electrons using a pitch angle-averaged kinetic equation allows
simulation times to be reduced almost to the level of pure fluid models while
the electron hot-tail generation is still accurately captured.

Together with the comprehensive physics model, which reaches beyond
previous efforts in kinetic disruption modelling, the code is equipped
with several attractive numerical features including fully implicit
time stepping of the full system, as well as a flux
conservative and positivity preserving discretization, which
contributes to the robustness of the tool. Due to its flexibility and
numerical efficiency, \DREAM\ is suitable for extensive investigations
of disruption and runaway physics.

\DREAM\ has been verified against both kinetic and fluid codes, and
has been found to reproduce their results in the appropriate limits. As
an application, two disruption scenarios were investigated in an
ASDEX Upgrade like tokamak, where the disruption is triggered by the
injection of a combination of neutral deuterium and argon atoms. The
full hierarchy of electron models was compared: {\em fully kinetic},
{\em superthermal}, {\em isotropic} and {\em fluid} models, and reasonable
agreement was found. The difference in simulation time between the {\em fluid}
and {\em kinetic} simulations is more than two orders of magnitude.
  
\section*{Acknowledgements}
The authors are grateful to Joan Decker and Yves Peysson for compiling
the \LUKE\ manual, and to S.~Newton, I.~Pusztai, and the rest of the Chalmers
Plasma Theory group for fruitful discussions. This work was supported by the
European Research Council (ERC) under the European Union’s Horizon
2020 research and innovation programme (ERC-2014-CoG grant 647121) and
the Swedish Research Council (Dnr.~2018-03911).

\bibliographystyle{elsarticle-num}
\bibliography{ref.bib}

\appendix
\section{Magnetic field and evaluation of flux surface integrals}\label{app:geometry}
\input{appGeometry.tex}

\section{Kinetic equation}\label{app:kineq}
\input{appKinEq.tex}

\section{Runaway fluid formulae in general tokamak geometry}\label{app:refluid}
\input{appRunawayFluid.tex}

\end{document}

%% file: defs.tex
\newcommand{\CODE}{\textsc{Code}}
\newcommand{\DREAM}{\textsc{Dream}}

\newcommand{\GO}{\textsc{Go}}
\newcommand{\LUKE}{\textsc{Luke}}

\newcommand{\Bmin}{\ensuremath{B_\mathrm{min}}}
\newcommand{\dd}{\mathrm{d}}
\newcommand{\ee}{\mathrm{e}}
\newcommand{\Ec}{E_{\rm c}}
\newcommand{\ED}{E_{\rm D}}
\newcommand{\Eceff}{\ensuremath{E_{\rm c}^{\rm eff}}}
\newcommand{\Epar}{E_\parallel}
\newcommand{\Jac}{\mathsf{J}}
\newcommand{\Mat}{\mathsf{M}}

\newcommand{\Rm}{R_{\rm m}}
\newcommand{\Vp}{\ensuremath{\mathcal{V}'}}
\newcommand{\VpVol}{\ensuremath{V'}}
\newcommand{\Zeff}{\ensuremath{Z_\mathrm{eff}}}

\newcommand{\ncold}{\ensuremath{n_\mathrm{cold}}}
\newcommand{\Tcold}{\ensuremath{T_\mathrm{cold}}}
\newcommand{\fhot}{\ensuremath{f_\mathrm{hot}}}
\newcommand{\nhot}{\ensuremath{n_\mathrm{hot}}}
\newcommand{\jhot}{\ensuremath{j_\mathrm{hot}}}
\newcommand{\fRE}{\ensuremath{f_\mathrm{re}}}
\newcommand{\nRE}{\ensuremath{n_\mathrm{re}}}
\newcommand{\jRE}{\ensuremath{j_\mathrm{re}}}
\newcommand{\pRE}{\ensuremath{p_\mathrm{re}}}
\newcommand{\EdotB}{\ensuremath{\left\langle \boldsymbol{E}\cdot\boldsymbol{B} \right\rangle }}

\newcommand{\FSA}[1]{\ensuremath{\left\langle #1 \right\rangle }}
\newcommand{\BA}[1]{\ensuremath{\left\{ #1 \right\}}}
\newcommand{\hypot}[2]{\ensuremath{\sqrt{ #1^2 + #2^2 }}}
\newcommand{\bb}[1]{\boldsymbol{#1}}

%% file: theory.tex
Electrons in \DREAM\ are primarily modelled with a bounce-averaged drift-kinetic equation
of the form
\begin{equation}\label{eq:transport}
    \begin{gathered}
        \frac{\partial f}{\partial t} =
            \sum_{m,n}
            \frac{1}{\Vp}\frac{\partial}{\partial z^m} \Bigg[
                \Vp\Bigg(
                    -\BA{A^m}f \\
                    +\BA{D^{mn}}\frac{\partial f}{\partial z^n}
                \Bigg)
            \Bigg]
            + \BA{S},
    \end{gathered}
\end{equation}
where $f(t,\,\boldsymbol{z})$ denotes the electron
distribution, $z^i$ denotes phase space coordinates and the contravariant components
$A^m = (\partial z^m/\partial \boldsymbol{z}) \cdot \boldsymbol{A}$ and (in dyadic notation)
$D^{mn} = (\partial z^m/\partial \boldsymbol{z}) (\partial z^n/\partial \boldsymbol{z}) : \mathsf{D}$,
where $\boldsymbol{A}$ and $\mathsf{D}$ represent the underlying advection vector and diffusion
tensor, driving electron phase-space flows.
We consider the zero-orbit-width limit, wherein the Larmor radius and cross
field drifts are neglected, so that electrons exactly follow magnetic field
lines, and choose $\bb{z}$ as the
constants of motion $\bb{z}=(r,p,\xi_0)$, where the flux-surface label $r$ is
the distance from the magnetic axis 
when the particle orbit passes the point of minimum magnetic field strength \Bmin; 
$\xi_0 = \boldsymbol{B}\cdot \boldsymbol{p}/(B p)|_{B=\Bmin}$ is the particle pitch with respect to the magnetic field $\bb{B}$ at this point and $p$ is the magnitude
of the particle momentum.
The kinetic equation~\eqref{eq:transport} has been averaged over the remaining three coordinates, using the \emph{bounce average} denoted with curly brackets 
\begin{align}
    \BA{X}&= \frac{1}{\Vp}\int_0^{2\pi}\mathrm{d}\zeta\int_0^{2\pi}\dd\phi\oint\dd\theta\sqrt{g} X, \nonumber \\
    \Vp &=\int_0^{2\pi}\mathrm{d}\zeta\int_0^{2\pi}\dd\phi\oint\dd\theta\sqrt{g},
\end{align}
with toroidal angle $\phi$, gyrophase $\zeta$ and where the angle $\theta$ is an arbitrary poloidal angle  which parametrizes the flux surface, assumed to be $2\pi$-periodic. The metric $\sqrt{g}$ of the $(r,\theta,\phi,p,\xi,\zeta)$ coordinates and the spatial Jacobian $\mathcal{J}$ are defined by
\begin{align}
    \sqrt{g} &= p^2\frac{B}{\Bmin}\frac{\xi_0}{\xi} \mathcal{J}, \nonumber \\
    \mathcal{J} &=  \frac{1}{|\nabla \phi \cdot (\nabla \theta \times \nabla r )|},
    \label{eq:jacobians}
\end{align}
where $\xi = \mathrm{sgn}(\xi_0)\sqrt{1-(1-\xi_0^2)B/\Bmin}$. 
The poloidal integral along the particle orbit is taken as
\begin{align}
    &\oint\dd\theta \, X(r,\,\theta,\,\phi,\,p,\,\xi,\,\zeta) \nonumber\\
    &= \begin{cases}
        \int_{-\pi}^{\pi}\dd\theta \, X(r,\,\theta,\,\phi,\,p,\,\xi,\,\zeta), \hspace{8.4mm} |\xi_0|>\xi_T  \\
        \int_{\theta_{b1}}^{\theta_{b2}}\dd\theta \, [X(r,\,\theta,\,\phi,\,p,\,\xi,\,\zeta)  \\ 
        \hspace{12mm} + X(r,\,\theta,\,\phi,\,p,\,-\xi,\,\zeta) ], \hspace{3mm} 0 < \xi_0 \leq \xi_T  \\
        0 , \hspace{39mm}  -\xi_T \leq \xi_0 \leq 0
    \end{cases}
    \label{eq:poloidalsplit}
\end{align}
where the bounce points $\theta_{b1}$ and $\theta_{b2}$ are the two separate poloidal angles 
defined by $\xi(\xi_0,\theta_{b1}) = \xi(\xi_0,\theta_{b2})=0$, and the trapped-passing boundary 
is denoted $\xi_T = \sqrt{1-\Bmin/B_\mathrm{max}}$. In the positive trapped region, $0<\xi_0\leq\xi_T$, the full contributions from both co-moving and counter-moving trapped particles are accounted for, and therefore all bounce averages are set to zero in $-\xi_T \leq \xi_0 \leq 0$ in order to avoid double counting (in this region the solution satisfies $f(\xi_0) = f(-\xi_0)$).
We will also utilize the spatial flux surface average 
\begin{align} \label{eq:FSA}
\FSA{Y} &= \frac{1}{V'}\int_0^{2\pi}\dd\phi \int_{-\pi}^{\pi} \dd \theta \, \mathcal{J}Y, \\
V' &= \int_0^{2\pi}\dd\phi \int_{-\pi}^{\pi} \dd \theta \, \mathcal{J}.
\end{align} 
The geometry of the magnetic surfaces enters into $\mathcal{J}$ and into the 
dependence on poloidal angle of $B(\theta)$; in \DREAM{} we evolve the equations
of motion in a static magnetic geometry where each flux surface is parametrized
by its major radius, elongation, Shafranov shift, triangularity and a
reference poloidal flux profile, as described in~\ref{app:geometry}. The
magnetic field is represented by the mixed form
\begin{equation}\label{eq:magfield}
\boldsymbol{B} = G \nabla \phi + \frac{1}{2\pi}\nabla \phi \times \nabla \psi,
\end{equation}
with $\psi$ the poloidal flux---labeling magnetic surfaces---defined as the magnetic flux through a horizontal disc centred on the tokamak axis of symmetry, as in Ref.~\cite{Boozer2005}.

\subsection{Kinetic equation}\label{sec:theory:kinetic}
\DREAM\ implements the following model for the phase-space advection and diffusion:
\begin{align}
\bb{A} &= \bb{A}_E + \bb{A}_C + \bb{A}_B + \bb{A}_S + \bb{A}_T, \\
\bb{D} &= \mathsf{D}_C + \mathsf{D}_T,
\end{align}
where $\bb{A}_E$ describes acceleration in an electric field, $\bb{A}_C$ collisional friction, $\bb{A}_B$ the bremsstrahlung radiation reaction force, $\bb{A}_S$ the synchrotron radiation reaction force and $\bb{A}_T$ radial transport. Similarly, $\mathsf{D}_C$ denotes collisional momentum-space diffusion and $\mathsf{D}_T$ radial diffusion. The particle source is modelled as
\begin{align}
S = C_\mathrm{ava} + S_p,
\end{align}
where $C_\mathrm{ava}$ denotes the knock-on collision operator responsible for the runaway avalanche, and $S_p$ a particle source to model electron density variations, for example due to ion transport or ionization-recombination.

In the following, we give explicit expressions for these terms in the bounce-averaged
electron drift-kinetic equation, as implemented in \DREAM. For the remainder of the paper,
$p = \gamma v/c$ denotes the relativistic momentum normalized to $m_e c$, with $\gamma = (1-v^2/c^2)^{-1/2}$ denoting the Lorentz factor and $v$ the speed. Other quantities are given in SI units, except temperatures which have the dimension of energy (the Boltzmann constant is $k_B=1$).

\subsubsection{Electric field}
The acceleration in the parallel electric field is described by the advection term 
\begin{equation}
\boldsymbol{A}_E = -e\boldsymbol{E},
\end{equation}
with the non-vanishing bounce-averaged components 
\begin{align}
\BA{A^p_E}&= -e\BA{E_\parallel \xi}, \nonumber \\
\BA{A^{\xi_0}_p} &= -e\frac{1-\xi_0^2}{p\xi_0} \BA{E_\parallel \xi},
\end{align}
where $E_\parallel = \boldsymbol{E}\cdot \boldsymbol{B}/B$, $e$ is the elementary charge, and the average can be written
\begin{align}
\hspace{-3mm} \BA{E_\parallel \xi} = 2\pi p^2 \xi_0 \frac{V'}{\Vp}\frac{\EdotB}{\Bmin}
\times \begin{cases}
0, & \text{trapped}, \\
1, & \text{passing}.
\end{cases}
\label{eq:app:kineq:BAE}
\end{align}

\subsubsection{Collision operator}
Collisions are modelled with a test-particle Fokker-Planck operator consisting of the advection and diffusion terms
\begin{align}
\mathsf{D}_C &= m_e \Tcold\gamma \nu_s \hat{p}\hat{p} + m_e^2c^2\frac{\nu_D}{2}p^2 (\mathsf{I} - \hat{p}\hat{p}), \nonumber \\
\boldsymbol{A}_C &= -\nu_s \boldsymbol{p},
\end{align}
with the momentum unit vector $\hat{p} = \boldsymbol{p}/p$. These have the non-vanishing 
components 
\begin{align}
\BA{D_C^{\xi_0 \xi_0}} &= (1-\xi_0^2)\frac{\nu_D}{2}\BA{\frac{\Bmin}{B}\frac{\xi^2}{\xi_0^2}}, \nonumber \\
\BA{D_C^{pp}} &=  m_e \Tcold\gamma \nu_s, \nonumber \\
\BA{A_C^p} &= -p\nu_s.
\end{align}
The collision frequencies $\nu_s$ and $\nu_D$ describe slowing down and pitch-angle scattering and, with the collision operator written in this form, support a Maxwell-J\"uttner equilibrium distribution at temperature \Tcold\ independently of their value. These test-particle collision frequencies are given by the sum of the contributions from different sources.

For collisions with free electrons, the collision frequencies are taken from the relativistic Coulomb Fokker-Planck operator~\cite{BeliaevBudker1956}
\begin{align}
\nu_s^\mathrm{ee} &= \nu_c\frac{\gamma^2 \Psi_1 - \Theta \Psi_0 + (\Theta\gamma-1) p e^{-(\gamma-1)/\Theta}}{p^3 e^{1/\Theta} K_2(1/\Theta)}, \nonumber \\
\nu_D^\mathrm{ee} &= \frac{\nu_c}{\gamma p^5 e^{1/\Theta}K_2(1/\Theta)}\biggl[ (p^2\gamma^2+\Theta^2)\Psi_0 + \Theta(2p^4-1)\Psi_1 \nonumber \\
&\hspace{20mm} + \gamma\Theta[1+\Theta(2p^2-1)]p e^{-(\gamma-1)\Theta}\biggr], \nonumber \\
\Psi_n &= \int_0^p  (1+s^2)^{(n-1)/2}e^{-(\sqrt{1+s^2} - 1)/\Theta}\, \mathrm{d}s,  \nonumber \\ 
\Theta &= \frac{\Tcold}{m_e c^2}, \nonumber \\
\nu_c &= 4\pi \ln\Lambda_{ee}\ncold r_0^2 c,
\label{eq:ee collfreq}
\end{align}
where $K_2$ is the second-order modified Bessel function of the second kind and $r_0 = e^2/(4\pi\varepsilon_0 m_e c^2)$ the classical electron radius. Here, \ncold{} and \Tcold{} refer to the density
and temperature of the Maxwellian component of the electron distribution, around which the collision operator has been linearized.

Ion collisions, assumed to be against infinitely massive targets, are accounted for using the model presented in Ref.~\cite{hesslow2018a}, yielding the contributions 
\begin{align}
\nu_D^\mathrm{ei} &= 4\pi \frac{\gamma}{p^3} c r_0^2 \sum_i n_i [\ln\Lambda_{ei} Z_{0i}^2 + g_i(p)], \nonumber \\
\nu_s^\mathrm{ei} &= 4\pi \frac{\gamma^2}{p^3} \sum_i n_i (Z_i-Z_{0i})\frac{1}{k}\ln(1+h_i^k), \nonumber\\
g_i &= \frac{2}{3}(Z_i^2 - Z_{0i}^2)\ln(1+(\bar{a}_i p)^{3/2}) \nonumber \\
& \hspace{10mm} - \frac{2}{3}(Z_i-Z_{0i})^2 \frac{\bar{a}_i p}{1+\bar{a}_i p}, \nonumber\\ 
h_i &= p\sqrt{\gamma-1}\frac{m_e c^2}{I_i}e^{-v^2/c^2},
\label{eq:screened-coll}
\end{align}
where the sum is taken over all ion species (and charge states) in the plasma, $Z_i$ denotes their atomic number and $Z_{0i}$ the charge number. The ad-hoc matching parameter $k=5$, and $\bar{a}_i$ is an effective-size parameter calculated and tabulated in Ref.~\cite{hesslow2018a}, or evaluated using the recommended formula $\bar{a}_i = (3/2\alpha)(\pi/3)^{1/3} (Z_i-Z_{0i})^{2/3}/Z_i$, with $\alpha \approx 1/137$ the fine-structure constant, in the absence of tabulated values. The ionic mean-excitation energy $I_i$ is calculated using tabulated values from Ref.~\cite{sauer2018} or extrapolated using their proposed 2-parameter formula\footnote{We adopt the values $D_N=D_{14}$ for $N=Z_i-Z_{0i}>14$ and $I_{0Z} = 10Z$ for the parameters $D_N$ and $I_{0Z}$ which appear in equation~(8) of Ref.~\cite{sauer2018}.} for $Z_i > 18$. The Coulomb logarithms are modelled using
\begin{align}
\ln\Lambda^\mathrm{ee} &= \ln\Lambda_0 + \frac{1}{k}\ln[1+[2(\gamma-1)/p_{Te}^2]^{k/2}], \nonumber \\
\ln\Lambda^\mathrm{ei} &= \ln\Lambda_0 + \frac{1}{k}\ln[1+(2p/p_{Te})^k], \\
\ln\Lambda_0 &= 14.9 + \ln\left(\frac{\Tcold}{1\,\text{keV}}\right) - 0.5\ln\left(\frac{n_\mathrm{free}}{10^{20}\,\text{m}^{-3}}\right),\nonumber 
\end{align}
with the free-electron density $n_\mathrm{free} = \sum_i Z_{0i}n_i$, $p_{Te} = \sqrt{2\Tcold/m_e c^2}$ is the normalized thermal momentum and $k=5$ is a matching parameter analogous to the one appearing in~\eqref{eq:screened-coll}.

\subsubsection{Radiation reaction force}
Radiation losses due to bremsstrahlung are captured using a mean-force model~\cite{KochMotz}
\begin{align}
\nu_s^B &= \frac{\gamma}{p} \alpha r_0^2 c \sum_i n_i Z_i^2 \Biggl[ \frac{12\gamma^2+4}{3\gamma p}\ln(\gamma+p) \nonumber \\
& - \frac{8\gamma + 6p}{3\gamma p^2}\ln^2(\gamma+p) - \frac{4}{3} + \frac{2F[2p(\gamma+p)]}{\gamma p}\Biggr], \nonumber \\
F(x) &= \int_0^x \frac{\ln(1+y)}{y}\mathrm{d}y, \\
\bb{A}_B &= -\nu_s^B\bb{p}, \nonumber
\end{align}
where effects of straggling~\cite{embreus2016} and screening are ignored.

Synchrotron emission due to the electron gyromotion around the magnetic field line is accounted for using the advection term
\begin{align}
\boldsymbol{A}_S &= -\frac{1}{\gamma \tau_S} \left[ \boldsymbol{p}_\perp + p_\perp^2 \boldsymbol{p}\right] , \nonumber \\
\frac{1}{\tau_S} &= \frac{e^4 B^2}{6\pi\varepsilon_0 m_e^3 c^3},
\end{align}
where $\boldsymbol{p}_\perp = \boldsymbol{p} - \boldsymbol{B}\cdot\boldsymbol{p}/B$ denotes the perpendicular momentum. This synchrotron advection term has 
the components
\begin{align}
\BA{A_S^p} &= \frac{1}{\tau_{S,\mathrm{min}}}p\gamma(1-\xi_0^2)\BA{\frac{B^3}{\Bmin^3}}, \nonumber \\
\BA{A_S^{\xi_0}} &= \frac{1}{\tau_{S,\mathrm{min}}}\frac{1}{\gamma}\xi_0(1-\xi_0^2)\BA{\frac{B^2}{B_\mathrm{min}^2}\frac{\xi^2}{\xi_0^2}},
\end{align}
where $\tau_{S,\mathrm{min}} = \tau_S(B=\Bmin)$.

\subsubsection{Avalanche source}
Avalanche generation is modelled using the Rosenbluth-Putvinski~\cite{RosenbluthPutvinski1997} source term
\begin{align}
\BA{C_\mathrm{ava}} &= \FSA{\nRE} n_\mathrm{tot}c r_0^2 \frac{\BA{B \delta(\xi - \xi^\star)}}{\FSA{B}} \frac{1}{p^2}\frac{\partial}{\partial p}\frac{1}{1-\gamma}, \label{eq:RP source} \nonumber \\
\xi^\star &= \sqrt{\frac{\gamma-1}{\gamma+1}},
\end{align}
where \nRE\ is the density of runaway electrons and $n_\mathrm{tot} = \sum_i Z_in_i$ denotes the total density of electrons (free and bound). The inclusion of bound electrons in this source term provides an approximation for the energy spectrum of electrons created via ionization, which is valid when the energy of the created electron is much greater than the binding energy of the ion~\cite{McDevitt2019a}.
The conservative discretization of this source is detailed in section~\ref{sec:disc-ava}.

\subsubsection{Radial transport}
Radial transport is captured by prescribing the contravariant radial coefficients $A^r = (\partial \bb{x}/\partial r)\cdot \bb{A}_T$ and $D^{rr} = (\partial \bb{x}/\partial r)(\partial \bb{x}/\partial r) : \mathsf{D}_T$. They can either be specified as arbitrary functions of $(t,\,r,\,p,\,\xi_0)$, or via a Rechester-Rosenbluth model representing diffusion in fully stochastic magnetic field line regions~\cite{rechester1978electron} 
\begin{equation} \label{eq:rechester-rosenbluth}
D^{rr} = \pi q \Rm \left(\frac{\delta B}{B}\right)^2 |v_\parallel| \mathcal{H}(\xi_0),
\end{equation}
where $(\delta B/B)(r,t)$ represents a normalized radial magnetic field fluctuation amplitude on the flux surface, $\Rm$ is the major radius of the magnetic axis, $q=q(r)$ is the safety factor calculated from the dynamically evolved plasma current density, and the step function $\mathcal{H}$ equals unity for passing particles and zero for trapped:
\begin{equation} \label{eq:H func}
\mathcal{H}(\xi_0) = \begin{cases}
1, & |\xi_0|>\xi_T \\
0, & |\xi_0|\leq \xi_T
\end{cases}
\end{equation}
with $\xi_T = \sqrt{1-\Bmin/B_\mathrm{max}}$ denoting the trapped-passing boundary. Since the Rechester-Rosenbluth model describes parallel transport along open field lines, trapped particles which bounce back and forth, will not undergo any net radial transport.
The bounce average of~\eqref{eq:rechester-rosenbluth} can concisely be formulated as $\BA{D^{rr}} = \pi q \Rm c(\delta B/B)^2 v|\xi_0|\BA{\xi/\xi_0}$.
The heat transport $D_W$ associated with this transport model is derived in~\ref{app:kineq:DW}.

\subsubsection{Particle source}
In order to describe a time-dependent electron density due to ionization dynamics, an ad-hoc source term is added, of the form
\begin{align}
\BA{S_p} = S_p \delta(\boldsymbol{p}),
\end{align}
where electrons are created (or removed) with zero kinetic energy in the ion rest frame. The same source term is used for replacing radially transported fast electrons with cold electrons, required for quasi-neutrality to be maintained. The source amplitude $S_p(t,r)$ is treated as an additional unknown quantity which is solved for non-linearly in the \DREAM\ equation system, under the constraint of quasineutrality
\begin{equation}
\int \frac{\Vp}{\VpVol}f\, \dd p \dd \xi_0 = \sum_i Z_{0i} n_i.
\end{equation}

\subsection{Three electron populations} \label{subsec:electrons}
\begin{figure}
    \centering
    \includegraphics[width=\columnwidth]{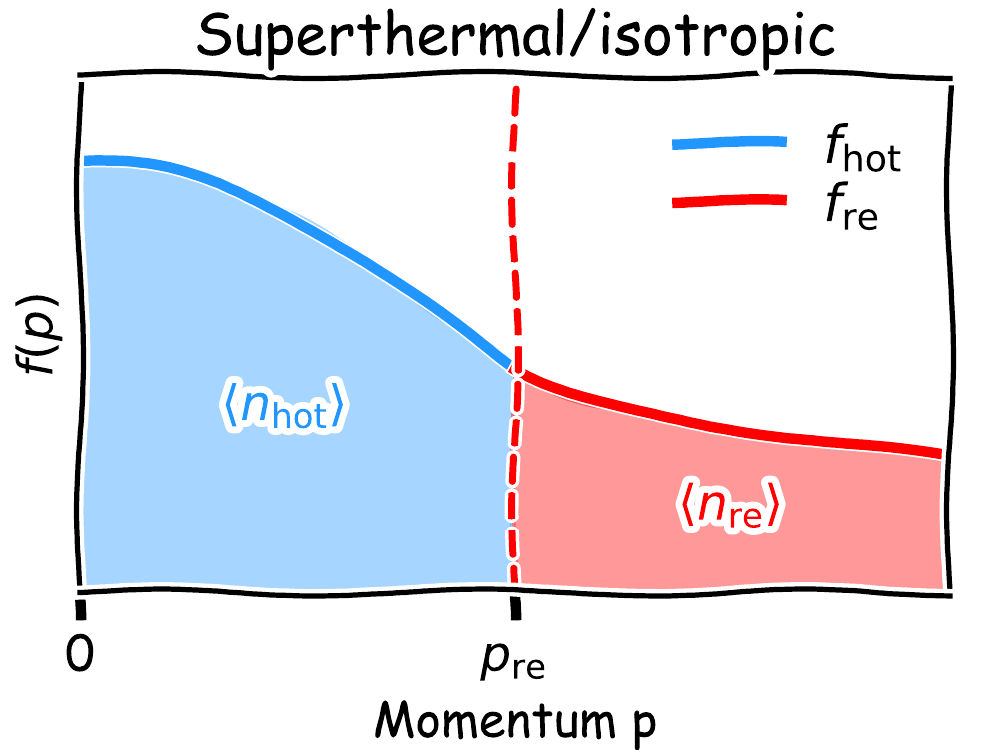}
    \caption{ \label{fig:f mode demo}
        Illustration of the three kinetic regions defined in the {\em fully
        kinetic} mode. In this mode, the distribution function $f_{\rm hot}$ describes both the
        \emph{cold} ($p<p_\mathrm{hot}$) and \emph{hot} ($p_\mathrm{hot}\leq p < p_\mathrm{re}$) electrons, while $f_{\rm re}$ describes the \emph{runaway} ($p \geq p_\mathrm{re}$) electrons on a separate simulation grid.
    }
\end{figure}

\DREAM\ allows the electron distribution to be evolved using the full kinetic equation~\eqref{eq:transport}, which is the most accurate but computationally expensive approach. The code also supports the solution of simplified equations for the electron dynamics which often provide adequately accurate results at significantly reduced computational cost. These approximations will be described in this section, and are based on the fact that electron dynamics are qualitatively different on three, typically well separated momentum scales; the corresponding characteristic momenta are
\begin{enumerate}
\item The thermal momentum $p_{Te} \sim 0.01$ at which the ohmic current, joule heating and many atomic-physics processes are dominant. 
\item The runaway critical momentum $p_c \sim \sqrt{E_c/E_\parallel} \sim 0.1$ at which the runaway generation rate is determined. In this region, electrons that will be thermalized are separated from those runaway electrons that are accelerated towards ultra-relativistic energies.
\item The acceleration region $p \sim \int eE_\parallel \, \dd t /m_e c \sim 100$ where the energy spectrum (and pitch distribution) of the runaway tail is set. The dynamics in this region determine the synchrotron and bremsstrahlung radiation emitted by a runaway beam, as well as the current decay rate during the runaway plateau. 
\end{enumerate}
In order to resolve each of these regions efficiently, as well as to allow flexibility in approximating parts of the electron dynamics (such as neglecting to resolve the energy spectrum of the runaway tail if only the generation rate is of interest), electrons in \DREAM\ are split into three separate populations: \emph{cold}, \emph{hot} and \emph{runaway} (\emph{re}) 
electrons which are, respectively, characterized by the densities \FSA\ncold, \FSA{n_\mathrm{hot}} and \FSA\nRE. These three momentum regions, as well as the two associated electron distribution functions \fhot\ and \fRE, are illustrated in figure~\ref{fig:f mode demo}. In the general case, these are distinguished only by labels on different phase-space regions, and the dynamics are purely governed by~\eqref{eq:transport} for all momenta. In the subsections that follow, we describe approximations to the electron kinetics in each of these three regions.

\subsubsection{Cold electrons}
The Maxwellian component of the electrons, the \emph{cold} population, is characterized by 
the density \FSA{\ncold}, temperature \Tcold{} and the parallel (ohmic) current density $j_\Omega$. The density represents the density of all free electrons that are not labelled \emph{hot} or \emph{runaway}:
\begin{equation} \label{eq:ncold}
\FSA\ncold = \FSA{n_\mathrm{free}} - \FSA{n_\mathrm{hot} }- \FSA{n_\mathrm{re}},
\end{equation}
where $\FSA{n_\mathrm{free}} = \sum_i Z_{0i} n_i$ is the total number density of free electrons, defined by the ion composition of the plasma, the sum taken over all ion charge states.
The temperature is evolved according to a transport equation, to be described further in section~\ref{sec:plasma transport}. 

The cold thermal component of the post-disruption plasma is not well-described by the bounce-averaged equation~\eqref{eq:transport} due to its typically high collisionality, and in addition the test-particle collision operator employed in \DREAM\ is not adequate for resolving the ohmic current due to the lack of field-particle collisions. Therefore, the ohmic current is modelled according to
\begin{equation}\label{eq:jomega}
\frac{j_\Omega}{B} = \sigma \frac{\EdotB}{\FSA{B^2}} + \frac{\delta j_\mathrm{corr}}{B},
\end{equation}
where $\sigma(n_i,\,\Tcold)$ denotes the parallel electric conductivity, and we implement the Sauter-Redl model~\cite{Redl2021} which accounts for neoclassical effects at arbitrary collisionality. The \emph{conductivity correction} $\delta j_\mathrm{corr}$ is an addition that is used to correct for transient currents when the thermal population is modelled kinetically (see section~\ref{sec:theory:hot} below), and is given by
\begin{align}\label{eq:corrcond}
\frac{\delta j_\mathrm{corr}}{B} &= -\frac{2\pi e}{\Bmin}\int_0^{p_\mathrm{hot}} \dd p \int_{-1}^1\dd \xi_0 \, p^2\mathcal{H}(\xi_0) v\xi_0 f_\mathrm{hot} \nonumber \\
&- \sigma_\mathrm{kineq}\frac{\EdotB}{\FSA{B^2}},
\end{align}
where the step function $\mathcal{H}$ is defined in~\eqref{eq:H func}, $p_\mathrm{hot}$ defines a threshold momentum (typically temperature dependent, $p_\mathrm{hot} = 7 p_{Te}$) separating cold from hot electrons, and $\sigma_\mathrm{kineq}$ represents the conductivity supported by the kinetic equation~\eqref{eq:transport} in a fully ionized plasma; in such plasmas, $\delta j_\mathrm{corr} = 0$ in steady state. The $\delta j_{\rm corr}$ term therefore provides a correction to the neoclassical conductivity in a rapidly time-varying plasma and accounts for the effect of partially ionized impurities.
By matching the Ohmic conductivity $\sigma_{\rm kineq}$ calculated with the test-particle collision operator for $Z_{\rm eff}\in[0,50]$ to the plasma conductivity calculated in Ref.~\cite{BraamsKarney}, it was found that the formula
\begin{align}
\sigma_\mathrm{kineq} = \left(1 - \frac{1.406}{1.888 + \Zeff}\right)\sigma_0
\end{align}
provides an accurate approximation, where $\sigma_0$ is the Sauter-Redl formula in the collisionless limit.

\subsubsection{Hot electrons}\label{sec:theory:hot}
The hot electrons, described by a distribution function \fhot, are modelled kinetically according to equation~\eqref{eq:transport}. They contribute density and parallel current density according to

\begin{align}\label{eq:nhot}
\FSA\nhot &= \int_{p_\mathrm{hot}}^{p_\mathrm{re}} \dd p \int_{-1}^1\dd \xi_0 \, \frac{\Vp}{\VpVol} \fhot, \\\label{eq:jhot}
\frac{\jhot}{B} &= -\frac{2\pi e}{\Bmin}\int_{p_\mathrm{hot}}^{p_\mathrm{re}} \dd p \int_{-1}^1\dd \xi_0 \, p^2\mathcal{H}(\xi_0) v\xi_0 \fhot,
\end{align}
where the two threshold momenta $p_\mathrm{hot}$ and $p_\mathrm{re}$ differentiate hot electrons from cold and runaway electrons, respectively. There are two main modes of treating hot electrons in \DREAM: the \emph{fully kinetic} mode where the full distribution function---including the thermal Maxwellian---is resolved kinetically on the grid, and \emph{superthermal} where only the non-Maxwellian part of the distribution is followed:

\paragraph{Fully kinetic} 
The full equation~\eqref{eq:transport} is solved, in the form presented in section~\ref{sec:theory:kinetic}.
Therefore, thermal electrons are resolved kinetically alongside hot electrons in the distribution function $f_{\rm hot}$ as illustrated in figure~\ref{fig:f mode demo}, since the collision operator is density conserving and drives the solution towards a Maxwell-J\"uttner distribution at temperature \Tcold. In this case, the \emph{cold} density is given by
\begin{equation}
\FSA\ncold = \int_{0}^{p_\mathrm{hot}} \dd p \int_{-1}^1\dd \xi_0 \, \frac{\Vp}{\VpVol} \fhot,
\end{equation}
corresponding to the density of electrons having momentum less than the hot-threshold $p_\mathrm{hot}$.

\begin{figure}
    \centering
    \includegraphics[width=\columnwidth]{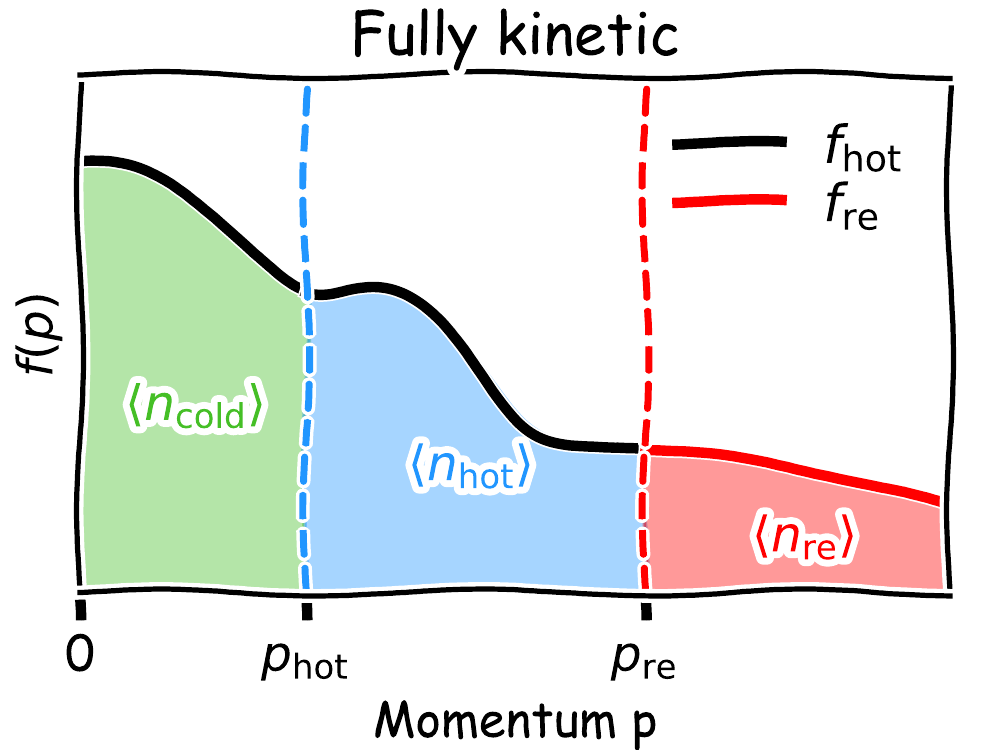}
    \caption{ \label{fig:f mode demo superthermal}
        Illustration of how the distribution functions $f_{\rm hot}$ and
        $f_{\rm re}$ are defined in the {\em superthermal} mode. In this mode,
        the hot electron distribution function $f_{\rm hot}$ covers only the
        \emph{hot} region which is extended to $p=0$, while \emph{cold} 
        electrons are modelled off the grid as a fluid of density \FSA{\ncold} 
        and ohmic parallel current density $j_\Omega/B$.
    }
\end{figure}

\paragraph{Superthermal}
Only the superthermal electrons are resolved by taking the electron-electron collision frequencies as their $\Tcold \to 0$ limits
\begin{align}
\nu_s^\mathrm{ee} &= \frac{\nu_c}{p v^2/c^2}, \nonumber \\
\nu_D^\mathrm{ee} &= \frac{\nu_c}{p^2 v/c},
\label{eq:superthermal-nu}
\end{align}
in which case the kinetic equation will no longer support a Maxwellian equilibrium distribution, but instead 
acquires a particle sink at $p=0$ since the advective particle flux $\lim_{p\to 0} \Vp p\nu_s$ is finite. In this case, the hot-electron threshold is chosen to be $p_\mathrm{hot} = 0$, meaning that all electrons in \fhot\ are considered hot, as illustrated in figure~\ref{fig:f mode demo superthermal}. By the cold-density equation~\eqref{eq:ncold}, the electrons that are lost at $p=0$ will be added to \FSA\ncold.
In this mode, the conductivity correction $\delta j_\mathrm{corr}$ in equation~\eqref{eq:corrcond} is dropped since ohmic current is not resolved kinetically.
This approach was pioneered in Ref.~\cite{Aleynikov2017} for thermal quench simulations.

\paragraph{Isotropic}
A further approximate form of the superthermal mode is supported by \DREAM, where 
the kinetic equation is analytically pitch-angle averaged based on an asymptotic expansion in which
pitch-angle scattering is assumed to dominate the electron dynamics. The procedure mirrors the approximations employed in the calculation of the avalanche growth rate in ref.~\cite{RosenbluthPutvinski1997},
but is generalized here to time-dependent situations which covers hot-tail formation.
Details on the derivation of this reduced kinetic equation are
provided in~\ref{app:kineq:reduced}, and the resulting expression is 
given by~\eqref{eq:app:transport}.

\begin{table*}
    \centering
    \caption{
        Summary of the approximations used in the four main operating modes of
        \DREAM. In the fluid, isotropic and superthermal modes, the cold
        electrons are represented with a fluid model (here indicated with the
        density $n_{\rm cold}$ and temperature $T_{\rm cold}$). Notably, the
        representation of the hot population varies between the models, being
        represented either as part of the same distribution function
        $f_{\rm hot}$ as the thermal electrons (fully kinetic), with a distribution
        function in the superthermal limit (superthermal), with an energy
        distribution in the superthermal limit (isotropic), or not at all
        (fluid).
    }
    \begin{tabular}{r|c|c}\toprule
        & {\bf Cold/thermal electrons} & {\bf Hot electrons} \\\midrule
                                        Fluid & $n_{\rm cold}, T_{\rm cold}$ & (not modelled)\\
        Isotropic & $n_{\rm cold}, T_{\rm cold}$ & $\int\dd\xi_0\,f_{\rm hot}$  ($v\gg v_{\rm th}$)\\
        Superthermal & $n_{\rm cold}, T_{\rm cold}$ & $f_{\rm hot}$ ($v\gg v_{\rm th}$)\\
        Fully kinetic & $f_{\rm hot}$ & $f_{\rm hot}$\\
        \bottomrule
    \end{tabular}
\end{table*}

\subsubsection{Runaway electrons}\label{sec:theory:re}
Runaways are described by the distribution function \fRE, which satisfies the same kinetic equation as the hot electrons \fhot, but is defined on a separate grid  spanning the interval $p\in[p_\mathrm{re},\,p_\mathrm{max}]$, where the runaway threshold is typically chosen as $p_\mathrm{re} \sim 1$, and the maximum resolved momentum is $p_\mathrm{max} \sim 100$. Note that the boundary $p_{\rm re}$ for the runaway grid is distinct from the usual critical momentum for runaway $p_{\rm c}$ sometimes used to define a runaway electron in other codes. The runaway boundary $p_{\rm re}$ used in \DREAM\ is constant in the simulation and must be appropriately chosen by the user so that both the hot and runaway electron distribution functions can be well approximated numerically. As long as the numerical grid is sufficiently resolved, the choice of the boundary $p_{\rm re}$ has no effect on the simulation results.

The reason for separating hot and runaway electrons is that the electron distribution in the hot generation region ($p\sim p_c$) is nearly isotropic, whereas in the runaway tail it can be extremely anisotropic. By utilizing a separate grid, resolution and discretization methods can be tailored to more efficiently resolve the runaway tail.
Although the runaway electron distribution couples relatively weakly to the plasma evolution since they move with parallel velocities near the speed of light, the runaway distribution is essential when coupling to synthetic radiation diagnostic tools~\cite{Hoppe_2020,Hoppe2021} or when considering kinetic instabilities driven by the runaways~\cite{BreizmanAleynikov2017Review,Liu_2021}.

Particle conservation is ensured by enforcing a boundary condition on \fRE\ at the lower $p$ boundary
\begin{align}\label{eq:re bc}
F_{\mathrm{re}} &= F_\mathrm{hot} + \left(\frac{\partial \FSA\nRE}{\partial t}\right)_\mathrm{fluid}, \nonumber \\
F_\mathrm{hot/re} &=\\
\int \Vp \Bigg(&\BA{A^p} f_\mathrm{hot/re} + \BA{D^{pp}}\frac{\partial f_\mathrm{hot/re}}{\partial p} \Bigg)\,\dd\xi_0 \Bigg|_{p=p_\mathrm{re}} \nonumber
\end{align}
using a method described in section~\ref{subsec:hot-re bc}, where $F_\mathrm{hot/re}$ denotes the particle flux through the shared $p=p_\mathrm{re}$ boundary on the respective kinetic grids, and $(\partial \FSA\nRE/\partial t)_\mathrm{fluid}$ denotes the runaway generation rate due to sources that are not modelled kinetically on the \emph{hot} grid. An example of such a source is Dreicer generation when using the \emph{superthermal} mode, in which case the thermal Maxwellian is not resolved so that no electrons are available for Dreicer acceleration.
The runaway density, corresponding to the number density of \fRE, therefore evolves according to
\begin{align}\label{eq:re evolution}
\frac{\partial \FSA\nRE}{\partial t} &= F_\mathrm{hot} + \left(\frac{\partial \FSA\nRE}{\partial t}\right)_\mathrm{fluid} \nonumber \\
&+ \frac{1}{V'}\frac{\partial}{\partial r}\left[V'\left( A^r_\mathrm{re} \FSA\nRE + D^{rr}_\mathrm{re}\frac{\partial \FSA\nRE}{\partial r}\right)\right].
\end{align}
If the runaway electron distribution function is solved for, the transport term in equation~\eqref{eq:re evolution} is replaced with a source term which is the momentum-space integral of the transport term in the kinetic equation~\eqref{eq:transport}:
\begin{equation}
    \begin{gathered}
        \left(\frac{\partial\FSA\nRE}{\partial t}\right)_{\rm transport} =
        \frac{1}{V'}\frac{\partial}{\partial r}\Bigg[V'\\
        \times\int\dd p\dd\xi_0\,\Vp\,\left( A^r \fRE + D^{rr}\frac{\partial \fRE}{\partial r}\right)\Bigg].
    \end{gathered}
\end{equation}
In case the runaway electron distribution function is not explicitly evolved,
the transport coefficients can instead be directly prescribed as functions of
only radius, or be integrated over momentum space in the manner described in
Ref.~\cite{Svensson2021}.
The parallel runaway current density is defined as
\begin{equation}
\frac{\jRE}{B} = -\frac{2\pi e}{\Bmin}\int_{p_\mathrm{re}}^{p_\mathrm{max}} \dd p \int_{-1}^1\dd \xi_0 \, p^2\mathcal{H}(\xi_0) v\xi_0 \fRE.
\end{equation}

The fluid runaway rate $(\partial \FSA\nRE/\partial t)_\mathrm{fluid}$ depends on the kinetic model employed, and support is implemented for modelling runaway generation due to Dreicer, hot-tail, Compton, tritium-decay and avalanche~\cite{Vallhagen2020,MartinSolis2017} depending on which equation terms are enabled. The fluid runaway models are described in~\ref{app:refluid}. If the \emph{hot} grid is disabled altogether, the runaway distribution \fRE\ can still be evolved using fluid models for the generation via the boundary condition~\eqref{eq:re bc}. Likewise, if the \emph{runaway} grid is disabled, runaway generation is still captured via~\eqref{eq:re evolution} where transport coefficients $A_\mathrm{re}$ and $D_\mathrm{re}$ can be imposed. In this case, the runaway current is instead determined by
\begin{align}
\frac{\jRE}{B} = \frac{ec \FSA\nRE}{\FSA{B}}.
\end{align}
If neither hot or runaway electrons are resolved kinetically, a fully fluid-like system is obtained,
corresponding approximately to the model contained in the \GO\ code \cite{Vallhagen2020,Feher2011,elongation}, generalized here to account for effects of arbitrary axisymmetric toroidal geometry.

%% file: backgroundplasma.tex
\DREAM\ utilizes a test-particle collision operator for the electron dynamics,
which refers to the Maxwellian electron component of the plasma as well as the
ion composition. In order to close the equation system, equations governing the
evolution of these quantities as well as the electric field must be introduced.
In this section we describe the evolution of the background plasma and how it
couples to the electrons.

Most equations for the background plasma can be written in the one-dimensional
transport equation form
\begin{equation}\label{eq:fluidtransp}
    \begin{gathered}
        \frac{\partial X}{\partial t} = \frac{1}{V'}\frac{\partial}{\partial r}\left[
            V'\left(
                -\FSA{A^{r}}X +
                \FSA{D^{rr}}\frac{\partial X}{\partial r}
            \right)
        \right]\\
        + \FSA{S},
    \end{gathered}
\end{equation}
which has a structure very similar to the kinetic
equation~\eqref{eq:transport}, but only evolves the quantity $X$ in the radial
coordinate $r$. In addition, it has the spatial Jacobian $V'$ instead of the full phase-space
Jacobian \Vp, and uses flux surface averages for its coefficients, defined by~\eqref{eq:FSA}, instead of
bounce averages. The similar structure is utilized when implementing
the equations, as the discretized forms of the equations are near identical with
only coefficients differing.

\subsection{Ions}
Ions are modelled by the densities $n_i^{(j)}$ of each species $i$ with atomic number $Z_i$, and charge state $j$ with charge number $Z_{0j}$, which are assumed to be uniformly distributed on flux surfaces. They are evolved according to
\begin{align}
\frac{\partial n_i^{(j)}}{\partial t} &= \left(I_i^{(j-1)} \FSA{\ncold} + \FSA{\sigma_{\mathrm{ion},i}^{(j-1)} v}\right)n_i^{(j-1)}  \nonumber \\
&-  \left(I_i^{(j)} \FSA{\ncold} + \FSA{\sigma_{\mathrm{ion},i}^{(j)} v}\right)n_i^{(j)}  \nonumber \\
& + R_i^{(j+1)}\FSA{\ncold} n_i^{(j+1)} - R_i^{(j)}\FSA{\ncold} n_i^{(j)},   \label{eq:ionrateequation}
\end{align}
where $I$ and $R$ denote ionization and recombination rate coefficients, respectively, which are extracted from the OpenADAS database~\cite{ADAS}. The kinetic ionization rates are modelled by
\begin{align}\label{eq:kineticioniz}
\FSA{\sigma_{\mathrm{ion},i}^{(j)} v}= \int  \mathrm{d}p\int_{-1}^1 \mathrm{d}\xi_0\,\frac{\Vp}{V'} v \sigma_{\mathrm{ion},i}^{(j)}  f_\mathrm{hot/re},
\end{align}
where the ionization cross-sections $\sigma_{\mathrm{ion},i}^{(j)}$ are taken as in Ref.~\cite{garland2020}, which extended the validity of the Burgess-Chidichimo model~\cite{burgess1983} to relativistic energies, with the momentum integration limits taken as in section~\ref{subsec:electrons}. Unlike the cited studies where model parameters were chosen from atomic data, \DREAM{} uses parameters which have been fitted to the OpenADAS ionization coefficients in the low-density (coronal) limit when the distribution is taken to be a Maxwellian  for a range of temperatures, assuming that a single shell dominates the ionization. This method provides a smooth transition between kinetic and thermal (fluid) ionization rate coefficients, improving numerical stability of the solver. If the hot electrons \fhot\ or the runaways \fRE\ are not resolved kinetically, the contribution to the ionization rate from such electrons is neglected.

\subsection{Temperature}
The background electron temperature \Tcold, which is assumed to be uniform on flux surfaces, is modelled via the evolution of the thermal energy $W_\mathrm{cold} = 3\FSA{\ncold}\Tcold/2$. The time evolution is governed by
\begin{align}\label{eq:heat}
\frac{\partial W_\mathrm{cold}}{\partial t} &= \frac{j_\Omega}{B}\FSA{\boldsymbol{E}\cdot\boldsymbol{B} }- \FSA{\ncold}\sum_i \sum_{j=0}^{Z_i-1}n_i^{(j)}L_i^{(j)}  \nonumber \\
&\hspace{-15mm} + \FSA{Q_c} + \frac{1}{V'}\frac{\partial}{\partial r}\left[V'\frac{3\FSA{\ncold}}{2}  \left(A_W \Tcold+ D_W\frac{\partial \Tcold}{\partial r}\right)\right],
\end{align}
where the advection coefficient $A_W$ and diffusion coefficient $D_W$ are either prescribed functions of time and radius, or derived from the particle transport model, as described in \ref{app:kineq}. The cold electron density has been pulled out of the radial derivative, consistent with the assumption that electrons cannot be transported independently of ions, in order to maintain quasineutrality. The collisional heat transfer to the cold population from hot and runaway electrons, as well as ions, is given by
\begin{align}
\FSA{Q_c} &= \int  \mathrm{d}p\int_{-1}^1 \mathrm{d}\xi_0\,\frac{\Vp}{V'} \Delta\dot{E}_{ee} f_\mathrm{hot/re} + \sum_i Q_{ei}, \nonumber \\
Q_{kl} &= \frac{\FSA{ nZ^2 }_k \FSA{nZ^2}_l e^4 \ln\Lambda_{kl}}{(2\pi)^{3/2}\epsilon_0^2 m_k m_l} \frac{T_k- T_l}{\left(\frac{T_k}{m_k}+\frac{T_l}{m_l}\right)^{3/2}}, \nonumber \\
\Delta \dot{E}_{ee} &= 4\pi \ncold r_0^2 \ln\Lambda_{ee} \frac{m_e c^4}{v},
\label{eq:collheat}
\end{align}
where we assume that hot and runaway electrons only deposit energy via elastic collisions with free cold electrons, thereby neglecting energy transfer by electron impact ionization.  The sum $i$ is taken over all ion species in the plasma, and it has been assumed that different charge states of the same ion species have the same temperature so that only the total density $n_i= \sum_j n_i^{(j)}$ and weighted charge $\FSA{ n Z^2 }_i = \sum_{j=0}^{Z_i} n_i^{(j)}Z_{0j}^2$ of each species $i$ appears. The ion temperature evolves according to the thermal-energy equation
\begin{equation}
    \frac{\partial W_i}{\partial t} = \sum_j Q_{ij} + Q_{ie},
\end{equation}
where $W_i = 3T_in_i/2$. The rate of energy loss $L_i^{(j)}$ by inelastic atomic processes is modelled via
\begin{align}
L_i^{(j)} = L_\mathrm{line} + L_\mathrm{free} + \Delta W_i^{(j)} (I_i^{(j)} - R_i^{(j)}),
\end{align}
where $L_\mathrm{line}$ is the radiated power by line radiation, $L_\mathrm{free}$ by recombination radiation and bremsstrahlung, and the last terms represent the change in potential energy due to excitation and recombination, with the rate coefficients as in equation~\eqref{eq:ionrateequation}. The ionization threshold $\Delta W_i^{(j)}$ is retrieved from the NIST database~\cite{NIST}, and the other rate coefficients are taken from OpenADAS.

\subsection{Current, electric field and poloidal flux}
The current is evolved via a mean-field equation for the poloidal flux~\cite{boozer2018pivotal}, 
\begin{align}
\frac{\partial \psi}{\partial t} &= -V_\mathrm{loop} + \frac{\partial}{\partial \psi_t}\left(\psi_t \mu_0 \Lambda \frac{\partial}{\partial \psi_t}\frac{j_\mathrm{tot}}{B}\right), \label{eq:pf} \\
\psi_t &= \frac{1}{2\pi}\int_0^r V'\FSA{\boldsymbol{B}\cdot\nabla\phi}\,\mathrm{d}r,
\end{align}
where the loop voltage is given by $V_\mathrm{loop} = 2\pi \FSA{\boldsymbol{E}\cdot\boldsymbol{B}}/\FSA{\boldsymbol{B}\cdot\nabla\phi}$, and the generalized Ohm's law for the system is given by $j_\mathrm{tot} = j_\Omega + j_\mathrm{hot} + j_\mathrm{re}$, where the total parallel current density $j_\mathrm{tot}$ satisfies Amp\`ere's law 
\begin{align}
2\pi\mu_0\FSA{\boldsymbol{B}\cdot\nabla\phi}\frac{j_\mathrm{tot}}{B} = \frac{1}{V'}\frac{\partial}{\partial r}\left[V'\FSA{\frac{|\nabla r|^2}{R^2}}\frac{\partial \psi}{\partial r}\right].
\end{align}
Note that $\psi$ here is distinct from the poloidal flux appearing in the
definition of the magnetic field~\eqref{eq:magfield}. The latter is taken as a static
parameter which specifies the magnetic geometry and enters into all bounce and flux surface
averages, while the $\psi$ of equation~\eqref{eq:pf} is a dynamically evolved
quantity that sets the current profile evolution. The magnetic equilibrium (i.e. shape of the flux surfaces) is therefore kept constant during a simulation.

The
second term on the right-hand-side of equation (\ref{eq:pf}), the \emph{hyperresistive} term, acts to flatten the current profile due to magnetic field line breaking, while conserving the helicity content of the plasma~\cite{boozer_1986}. The derivatives with respect to toroidal flux $\psi_t$ are taken at constant radius $r$. The helicity transport coefficient $\Lambda(t,\,r)$, which is allowed to be arbitrarily prescribed, can be estimated e.g.~by experimental observations of the temporal current spike and drop in the internal inductance in disruptions \cite{Boozer_2019}.

The boundary condition for the poloidal flux is modelled by introducing $\psi_\mathrm{edge} = \psi(r=a)$ and $\psi_\mathrm{wall} = \psi(r=b)$, where $a$ is the (low-field side mid-plane) minor radius of the plasma and $b$ represents the radius of the conducting wall. The flux at the edge couples to that at the wall via an approximate edge-wall mutual flux inductance $M_{we} = \mu_0 \Rm \ln\frac{b}{a}$ with $\Rm$ the major radius of the plasma, such that $\psi_\mathrm{edge} = \psi_\mathrm{wall} - M_{we} I_p$. The total toroidal plasma current is
\begin{equation}
I_p = \frac{1}{2\pi}\int_0^a V'\FSA{\boldsymbol{B}\cdot\nabla\phi}\frac{j_\mathrm{tot}}{B}\,\mathrm{d}r.
\end{equation}
The poloidal flux $\psi_\mathrm{wall}$ at the tokamak wall is evolved by 
\begin{equation}
\frac{\partial \psi_{\rm wall}}{\partial t} = V_\mathrm{loop}^\mathrm{(wall)},
\end{equation}
where the loop voltage $V_\mathrm{loop}^\mathrm{(wall)}$ at the tokamak wall 
can be modelled in two different ways, given as follows:

\paragraph{Poloidal flux boundary condition 1: Prescribed} The wall loop voltage is provided as a prescribed time-dependent input, and the initial condition for the wall poloidal flux is chosen as $\psi_\mathrm{wall}(t=0)=0$.

\paragraph{Poloidal flux boundary condition 2: Self-consistent with circuit equation} The wall loop voltage is modelled by assuming an external inductance $L_\mathrm{ext} = \mu_0 \Rm \ln(\Rm/b)$ and a wall resistivity $R_\mathrm{wall}$ such that the characteristic wall time is $\tau_\mathrm{wall} = L_\mathrm{ext} / R_\mathrm{wall}$. Then, the wall loop voltage is defined as 
\begin{equation}
V_\mathrm{loop}^\mathrm{(wall)} = R_\mathrm{wall} I_\mathrm{wall},
\end{equation}
where the wall current $I_\mathrm{wall}$ is given in terms of the external inductance as
\begin{equation}
\psi_\mathrm{wall} = - L_\mathrm{ext} (I_p + I_\mathrm{wall}).
\end{equation}

%% file: implementation.tex
The main part of the codebase constituting \DREAM\ is written in C++, with a
rich interface written in Python. The C++ code is divided into two libraries
that contain all routines necessary for setting up and running simulations, and
one executable, which merely provides a thin command-line interface to the
library routines. The difference between the two libraries is that one library
contains lower-level routines for building finite volume stencils and
interacting with the sparse linear algebra library
PETSc~\cite{petsc-user-ref,petsc-efficient}, while the other library contains
code for the physical models available, for initializing simulations, as well as
for evolving the equation system in time.

In this section we will describe the details of the implementation,
particularly the time, space and momentum discretizations used, but also the
special boundary condition used to connect the hot and runaway electron kinetic
grids. We end the section with a discussion about the performance of the code
as well as some special methods that are used to improve performance of the
code.

\subsection{Finite volume discretization}\label{sec:fvm}
To preserve the integral of the evolved quantities, we discretize
equation~\eqref{eq:transport} using a finite volume method~\cite{Karney1986}.
This involves dividing the computational grid into $N_r\times N_p\times N_\xi$
cells with center values denoted using integer indices $(i,j,k)$ (henceforth
referred to as the \emph{distribution grid}), and points on the cell faces
denoted with half integers, e.g.\ $(i-1/2,j,k)$ (referred to as the \emph{flux
grid}). In \DREAM, grids are defined by specifying the edge points $z^m_{1/2}$
and $z^m_{N_m+1/2}$ of each flux grid along with the cell widths $\Delta z^m_i$
so that the flux and distribution grid points are distributed according to
\begin{equation}
    \begin{aligned}
        z^m_{i+1/2} &= z^m_{i-1/2} + \Delta z^m_i,\\
        z^m_i &= \frac{z^m_{i+1/2} + z^m_{i-1/2}}{2}.
    \end{aligned}
\end{equation}
When evaluating second derivatives it is also useful to introduce the
distribution grid spacing
\begin{equation}
    \Delta z^m_{i+1/2} = z^m_{i+1}-z^m_i.
\end{equation}

Next, the Fokker--Planck and transport equations~\eqref{eq:transport}
and~\eqref{eq:fluidtransp}, are averaged separately over each cell volume,
resulting in a set of coupled equations for the average value of $X$ in each
cell. By using a central difference approximation for phase-space derivatives,
and an Euler backward scheme for time derivatives, the discretized form of
equation~\eqref{eq:transport} becomes
\begin{equation}\label{eq:disc:transport}
    \begin{gathered}
        \frac{X^{(l+1)}_{ijk} - X^{(l)}_{ijk}}{\Delta t^{(l)}} =\\
            \frac{1}{\Vp_{ijk}\Delta z^m_{ijk}}\Bigg[
                            \Vp_{i_m+1/2}\left\{A^{m,(l+1)}\right\}_{i_m+1/2} X^{(l+1)}_{i_m+1/2} -\\
                \Vp_{i_m-1/2}\left\{A^{m,(l+1)}\right\}_{i_m-1/2}X^{(l+1)}_{i_m-1/2} +\\
                            \Vp_{i_m+1/2}\left\{D^{mn,(l+1)}\right\}_{i_m+1/2}\left.\frac{\partial X}{\partial z^n}\right|^{(l+1)}_{i_m+1/2} -\\
                \Vp_{i_m-1/2}\left\{D^{mn,(l+1)}\right\}_{i_m-1/2}\left.\frac{\partial X}{\partial z^n}\right|^{(l+1)}_{i_m-1/2}
            \Bigg] +\\
                        \left\{S\right\}^{(l+1)}_{ijk},
    \end{gathered}
\end{equation}
where $l$ is the time step index and the notation $i_m\pm N$ indicates that $N$
should be added to/subtracted from the index of the $m$'th phase space
coordinate, $z^m$. The discretized form of equation~\eqref{eq:fluidtransp} is
almost identical. Since the advection and diffusion fluxes are evaluated on
cell faces, the flux into any given cell on the computational grid will also be
exactly the flux out of adjacent grid cells. This guarantees exact conservation
of the integral of the quantity $X$ within machine precision, in the absence of
sources and edge losses.

\subsubsection{Discretization of advection terms}\label{sec:discradv}
The main difficulty of evaluating advection terms comes from the fact that the
quantity $X$ must be evaluated on a cell face rather than in the center of the
cell, where it is actually computed. To do so we must interpolate in $X$ based
on its value in adjacent cells and many possible interpolation schemes could be
used to this end. However, choosing the interpolation scheme with care can
provide benefits such as improved stability of the numerical scheme as well as
the preservation of monotonicity in $X$. In \DREAM, we generally let
\begin{equation} \label{eq:adv interp}
    X_{i_m-1/2} = \sum_{k=-2}^1 \delta^{(i_m)}_k X_{i_m+k},
\end{equation}
with $\left\{ \delta^{(i_m)} \right\}_k$ denoting a set of interpolation
coefficients to be determined. The user can then select among a range of popular
schemes, including linear schemes such as the simple centred ($\delta_{-1}=\delta_0=1/2$, for uniform grids), 
first or second-order upwind schemes, the third-order quadratic upwind scheme~\cite{Leonard1979},
or nonlinear flux limiter schemes such as SMART~\cite{Gaskell1988},
MUSCL~\cite{VanLeer1979}, OSPRE~\cite{Waterson1995} or TCDF~\cite{Zhang2015}.
The nonlinear schemes are designed to preserve positivity of the solution.
The flux limited schemes are upwind-biased, and for positive flow ($A_{i_m-1/2} \geq 0$) can 
be expressed as
\begin{align}
\delta_{-2} &= -k \phi(r), \nonumber\\
\delta_{-1} &= 1 + k\phi(r) 
\label{eq:flux limiter} \\
k &= \frac{x_{i_m-1/2} - x_{i_m-1}}{x_{i_m-1}-x_{i_m-2}}, \nonumber\\
r &= \frac{(X_{i_m}-X_{i_m-1})/(x_{i_m}-x_{i_m-1})}{(X_{i_m-1}-X_{i_m-2})/(x_{i_m-1}-x_{i_m-2})}.\nonumber
\end{align}
The flux limiter function $\phi(r)$ determines which scheme is used. 
For negative flows ($A_{i_m-1/2} < 0$), equation~\eqref{eq:flux limiter} is 
modified by mirroring all indices according to $i_m + x \mapsto i_m - 1 - x$
and $\delta_k \mapsto \delta_{-1-k}$.
The OSPRE and TCDF limiters prescribe continuously differentiable $\phi$, making them 
robust choices with attractive convergence properties in the Newton solver, whereas 
the other piecewise linear schemes may sometimes fail to converge, but may provide higher accuracy.

\subsubsection{Discretization of diffusion terms}
The derivatives appearing in the diffusion terms in
equation~\eqref{eq:transport} are conveniently also discretized using a central
difference approximation. This causes the diagonal ($m=n$) terms to take the
form
\begin{equation}
    \left.\frac{\partial X}{\partial z^n}\right|_{i_m-1/2} =
        \frac{X_{i_n} - X_{i_n-1}}{\Delta z^n_{i_n-1/2}},
\end{equation}
which also ensures that monotonicity is preserved for $X$. Off-diagonal
diffusion terms require interpolation, and thus do not preserve monotonicity.
In \DREAM, they are generally discretized as
\begin{equation}
    \begin{gathered}
        \left.\frac{\partial X}{\partial z^n}\right|_{i_m-1/2} =\\
            \frac{X_{i_m+1,i_n} + X_{i_m+1,i_n-1} - X_{i_m-1,i_n} - X_{i_m-1,i_n-1}}
            {\Delta z^n_{i_n+1/2} + \Delta z^n_{i_n-1/2}}.
    \end{gathered}
\end{equation}
An alternative to the above manner for discretizing off-diagonal diffusion terms
was given in~\cite{duToit2018} where the terms were rewritten as advection terms
and combined with a flux limiter scheme to ensure the preservation of
monotonicity. 

\subsection{Discretization of avalanche source}\label{sec:disc-ava}
The cell average of the bounce-averaged avalanche source~\eqref{eq:RP source} is evaluated according to
\begin{align}
\hat{S}_{ij} &= \frac{1}{\mathcal{V}'_{ij}\Delta p_i \Delta \xi_j}\int_{p_{i-1/2}}^{p_{i+1/2}}\mathrm{d}p\int_{\xi_{j-1/2}}^{\xi_{j+1/2}}\mathrm{d}\xi_0 \, \mathcal{V}'\BA{C_\mathrm{ava}} \nonumber \\
&= \frac{e^4}{8\pi\varepsilon_0^2 m_e^2 c^3}\frac{2\pi V'}{\mathcal{V}'_{ij}\Delta p_i \Delta \xi_j} \frac{n_\mathrm{tot}\FSA{n_\mathrm{re}}}{\FSA{B/\Bmin}} \nonumber \\
&\times \left(\frac{1}{\gamma_{i-1/2}-1}-\frac{1}{\gamma_{i+1/2}-1}\right)\hat\delta_{ij}, \nonumber \\
  \hat\delta_{ij} &= \frac{1}{V'} \int_0^{2\pi}\mathrm{d}\phi \oint \mathrm{d}\theta \,\mathcal{J}\frac{B}{\Bmin},
                     \label{eq:ava}
\end{align}
where the $\theta$ integral is taken over all angles for which $\xi(\xi_{j-1/2}) < \xi^\star(p_i) < \xi(\xi_{j+1/2})$, with $\xi(\xi_0)$ defined as in equation~\eqref{eq:jacobians}, $\xi^\star$ as defined in equation~\eqref{eq:RP source}, and it has been assumed that $n_\mathrm{re}/B$ is constant on flux surfaces (consistent with the assumption that runaways have $\xi=1$ in the derivation of the local source function). Equation~\eqref{eq:ava} is equivalent to equation (20) in Ref.~\cite{Nilsson_2015}.
We introduce a cutoff $\gamma_\mathrm{cut}$ such that, if $\gamma_{i-1/2} < \gamma_\mathrm{cut} < \gamma_{i+1/2}$, we replace $\gamma_{i-1/2} = \gamma_\mathrm{cut}$, and we set the source to 0 for $\gamma_{i+1/2} < \gamma_\mathrm{cut}$.
Defined this way, the numerical integral of the source term is
\begin{align}
\sum_{ij}\frac{\mathcal{V}'_{ij}}{V'} \Delta p_i \Delta \xi_j \hat{S}_{ij} &= \frac{e^4}{8\pi\varepsilon_0^2m_e^2 c^3}n_\mathrm{tot}\FSA{n_\mathrm{re}}\nonumber \\
&\times\left(\frac{1}{\gamma_\mathrm{cut}-1}  - \frac{1}{\gamma_\mathrm{max}-1}\right),
\end{align}
with $\gamma_{\rm max} = \sqrt{1+p_{\rm max}^2}$, and $p_{\rm max}$ denoting
the maximum momentum resolved on the grid. This expression agrees with an exact
integration of~\eqref{eq:RP source}, independently of grid resolution.

\subsection{Time evolution}
As described in section~\ref{sec:fvm}, \DREAM\ 
uses an Euler backward (implicit) time discretization. All quantities on the
right hand side of~\eqref{eq:disc:transport} should therefore be evaluated at
time $t=t_{l+1}$, the same time for which a solution to the equation system is
sought, thus requiring the nonlinear system to be solved iteratively. In
\DREAM, a standard Newton's method is used. By moving all terms
in~\eqref{eq:disc:transport} to the same side of the equality we obtain a
system of coupled equations in homogeneous form which we denote with the
operator $\bb{F}$. By also denoting the vector of unknowns at time $t_{l+1}$ by
$\bb{x}^{(l+1)}$, we can write the system of equations compactly as
\begin{equation}\label{eq:nonlineareq}
    \bb{F}\left(\bb{x}^{(l+1)}\right) = 0.
\end{equation}
In Newton's method we then linearize $\bb{F}$ around the true solution and
obtain the iterative scheme
\begin{equation}\label{eq:newton}
    \bb{x}^{(l+1)}_{i+1} = \bb{x}^{(l+1)}_i - \Jac^{-1}\left( \bb{x}^{(l+1)}_i \right)
        \bb{F}\left( \bb{x}^{(l+1)}_i \right),
\end{equation}
where the index $i$ indicates the Newton iteration and $\Jac^{-1}$ denotes the
inverse of the Jacobian of $\bb{F}$. The Jacobian
is constructed based on analytical expressions for all equations, although some
derivatives are approximated or even neglected altogether for simplicity and to
ensure the sparsity of the Jacobian matrix. Notably, since the kinetic equation
is relatively insensitive to the ion charge state distribution, a net performance gain
is typically obtained by neglecting the ion Jacobian in the kinetic equation. Also, the advection 
interpolation coefficients~\eqref{eq:adv interp} can be set to a linear \emph{upwind} 
scheme in the Jacobian, even if a flux limiter method is used in the evaluation of the 
residual $\bb{F}$; the computational gain by reducing the number of non-zeros in the Jacobian
can sometimes offset the additional iterations needed by the Newton solver to reach 
convergence.

An alternative approach to solving~\eqref{eq:nonlineareq}, suitable for linear
systems and nonlinear systems which vary slowly in time, is obtained with a
so-called \emph{linearly implicit} or \emph{semi-implicit} time discretization.
With this scheme, the equation system~\eqref{eq:disc:transport} is linearized in
time so that it can be written
\begin{equation}
    \bb{F}\left(\bb{x}^{(l+1)}\right)\approx \Mat\left( \bb{x}^{(l)} \right)\bb{x}^{(l+1)}
    + \bb{S}\left(\bb{x}^{(l)}\right),
\end{equation}
where $\Mat(\bb{x}^{(l)})$ is a matrix operator representing the differential
operators and coefficients in~\eqref{eq:disc:transport}, and
$\bb{S}(\bb{x}^{(l)})$ is a vector representing the sources $S$ in the same
equation. In contrast to the Newton's method described above, $\Mat$ and
$\bb{S}$ are here to be evaluated at the current time $t_l$ instead of at the
time for which the solution $\bb{x}^{(l+1)}$ is sought. As a result, a solution
for equation~\eqref{eq:nonlineareq} can be obtained using only a single matrix
inversion, instead of a series of repeated inversions as required by the Newton
method.

Both the linear and nonlinear solver options are available for all combinations
of equations in \DREAM, allowing the user to easily switch between the two to
compare performance and accuracy. To avoid duplicating implementations, the
matrix $\Mat$ and the residual $\bb{F}$ are both built using the same routines,
which receive a function pointer that is used to set a single element of $\Mat$,
or add to a single row of the residual $\bb{F}$.

\subsection{Convergence condition}
The Newton iteration~\eqref{eq:newton} should be terminated once the solution
$\bb{x}^{(l+1)}_{i+1}$ is sufficiently close to the true solution
$\bb{x}^{(l+1)}_\star$. This is typically done by evaluating an appropriate norm
of the Newton step
$\Delta\bb{x}^{(l+1)}_{i+1}=\bb{x}^{(l+1)}_{i+1}-\bb{x}^{(l+1)}_i$. This
method is applied in \DREAM\ as well, but since the unknown vector $\bb{x}$
is a combination of all the unknowns of the equation system, we calculate the
norm separately for each unknown quantity. To each unknown $X_n(r,p,\xi_0)$ we
assign a pair of absolute and relative tolerances, $\epsilon_n^{\rm abs}$ and
$\epsilon_n^{\rm rel}$ respectively, and demand that
\begin{equation}
    \left\lVert X^{(l+1)}_{n,i+1} - X^{(l+1)}_{n,i} \right\rVert_2 \leq
    \epsilon_n^{\rm abs} + \epsilon_n^{\rm rel}\left\lVert X^{(l+1)}_{n,i+1} \right\rVert_2,
\end{equation}
be satisfied for every unknown of the equation system $X_n$ in order for the
solution to be accepted. The norm is taken to be the 2-norm of the solution
vector corresponding to $X_n$.

By default, the relative tolerance is set to $\epsilon_n^{\rm rel} = 10^{-6}$
for all unknowns $X_n$. The absolute tolerance, on the other hand, is disabled
for most quantities (i.e.\ $\epsilon_n^{\rm abs} = 0$), with the exception of
the runaway density $n_{\rm RE}$ and current $j_{\rm RE}$, for which the absence
of an absolute tolerance can sometimes cause the Newton solver to
diverge\footnote{This is related to approximations in the Jacobian for
$n_{\rm RE}$ which causes the Newton solver to eventually oscillate around the
true solution $n_{\rm RE} = 0$ at a level which is physically insignificant.
}. We set the absolute tolerance for
$n_{\rm RE}$ to $\epsilon_{n_{\rm RE}}^{\rm abs} = 10^{-10}$ by default, and
for $j_{\rm RE}$ to
$\epsilon_{j_{\rm RE}}^{\rm abs} = ec\epsilon_{n_{\rm RE}}^{\rm abs}$.

\subsection{Treatment of fluid-kinetic boundary conditions}\label{subsec:hot-re bc}

One of the novel features of \DREAM\ is the ability to seamlessly evolve
different energy regions of the electron population using either fluid or
kinetic models. The connection between the different energy regions is handled
using specialized boundary conditions on the kinetic grids, and source terms on
the fluid grids. Two particularly interesting use cases are when a runaway
electron distribution function is included in an otherwise pure fluid
simulation, as well as the case where two separate distribution functions are
used to model the hot and runaway regions of the electron population.

\subsubsection{Fluid runaway sources on kinetic grids}
The use of a separate distribution function $f_{\rm RE}$ for the runaway region
in \DREAM\ makes it easy to resolve the high-energy runaway electron
distribution function in simulations which otherwise only evolve fluid
quantities. In such simulations, the electron population is split into a (free)
thermal electron density $n_{\rm cold}$ and a runaway density $n_{\rm RE}$.
The production of runaway electrons is modelled using fluid generation rates,
according to equation~\eqref{eq:re evolution}, with $F_{\rm hot}=0$, which move particles from the thermal
population to the runaway population.

When the runaway electron distribution function is included in such a
simulation, the particles that are moved to the runaway density $n_{\rm RE}$
must also be introduced to the runaway distribution function $f_{\rm RE}$. In
reality, the details of how the particles should be introduced depend on the
physics of the runaway production mechanism, but for the purpose of obtaining a
lightweight fluid-kinetic model we introduce particles in the cell corresponding
to $(p,\xi_0) = (p_{\rm min}, \mathrm{sign}(E))$. This choice
of source term is motivated by the fact that the newly introduced runaway
electrons will generally be rapidly accelerated to even higher energies in the
runaway electron distribution function, and thus move close to the $\xi_0=\pm 1$
boundary anyway. The final form of the distribution function is generally
\emph{not} set by the kinetic details of the runaway source term, but rather
by effects such as pitch-angle scattering, electric field acceleration,
synchrotron emission etc.\ which are all included fully kinetically in this
simulation mode. It is only the kinetic details of how particles cross into the
runaway region which are not resolved in this mode.

\subsubsection{Flow between hot and runaway distributions}
The separation of the electron distribution function into a hot region and a
runaway region can have significant computational advantages. The hot electron
distribution function is often close to isotropic while varying rapidly with
momentum, thus requiring only a few grid points in pitch $\xi_0$, but
potentially a large number of grid points in momentum $p$. The runaway electron
distribution function, on the other hand, is typically aligned close to
$\xi_0=\mathrm{sign}(E)$ with a long, slowly varying tail in momentum. It thus
often requires careful placement of pitch grid points, while only a handful of
grid points in momentum may be necessary.

\begin{figure}
    \centering
    \includegraphics[width=0.5\textwidth]{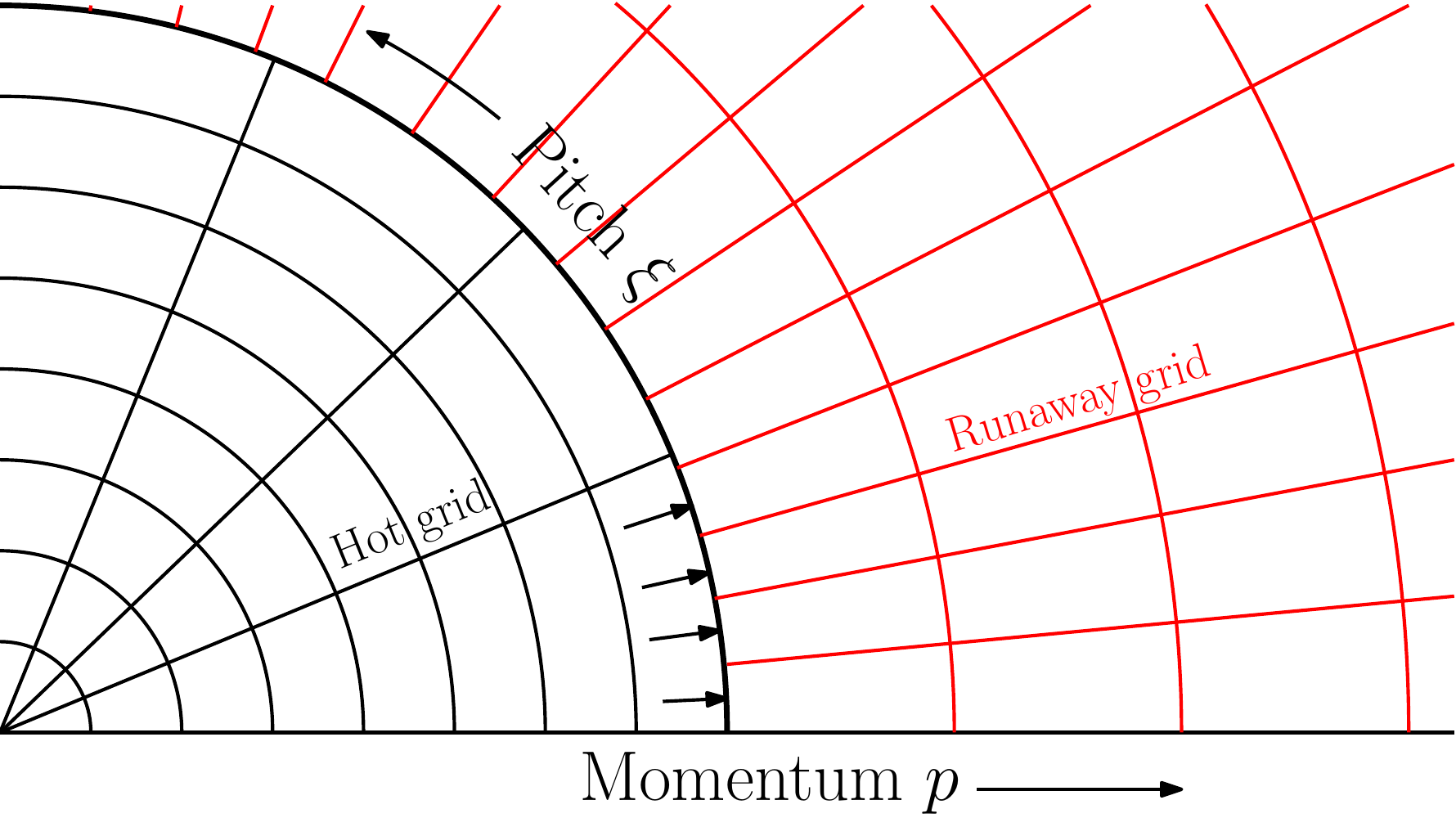}
    \caption{
        The hot (black) and runaway electron (red) kinetic grids are independent
        of each other but should allow for particles to flow freely between
        them.
    }
    \label{fig:kinetickineticgrid}
\end{figure}

To handle the flow of particles between the hot and runaway electron
distribution functions, a boundary condition connecting grid cells on both
sides of the interface between them is introduced, as illustrated in
figure~\ref{fig:kinetickineticgrid}. Since the two distribution functions may
be defined on grids with widely different resolutions, care must be taken to
ensure that the number of particles are conserved when electrons enter the
runaway grid and vice versa. The boundary condition connecting the two grids is
therefore based on the local density conservation equation
\begin{equation}\label{eq:localflux}
    \begin{gathered}
        \Phi^{(p),\rm RE}\left(p^{\rm RE}_{1/2},\xi^{\rm RE}_J\right)
        \Vp^{\rm RE}_{1/2,J} \Delta\xi^{\rm RE}_J =\\
        -\sum_j \Phi^{(p),\rm hot}\left( p^{\rm hot}_{\rm max}, \xi^{\rm hot}_j \right)
        \Vp^{\rm hot}_{N_p+1/2,j}\overline{\Delta\xi}_{jJ},
    \end{gathered}
\end{equation}
where $\Phi^{(p),\rm RE}$ and $\Phi^{(p),\rm hot}$ denote the particle fluxes
into the runaway grid and out of the hot grid, respectively, and the extent
$\overline{\Delta\xi}_{jJ}$ by which the cells overlap in $\xi_0$ is
\begin{equation}
    \begin{aligned}
        \overline{\Delta\xi}_{jJ} &= \min\left( \xi^{\rm hot}_{j+1/2}, \xi^{\rm RE}_{J+1/2} \right)\\
        &- \max\left( \xi^{\rm hot}_{j-1/2}, \xi^{\rm RE}_{J-1/2} \right).
    \end{aligned}
\end{equation}
By requiring the flux of particles to be locally conserved as
in~\eqref{eq:localflux}, one obtains for the advective and diffusive fluxes on
the hot electron grid
\begin{equation}\label{eq:kk:advdiffflux}
    \begin{gathered}
        \Phi^{(p),\rm hot}_{\rm adv}\left( p^{\rm hot}_{N_p+1/2}, \xi^{\rm hot}_j\right) =\\
        F_{N_p+1/2,j}^{\rm hot}f^{\rm hot}_{N_p+1/2,j} =\\
        F_{N_p+1/2,j}^{\rm hot}\left[\delta^{(1)}_j f^{\rm hot}_{N_p,j} + \left( 1-\delta^{(1)}_j \right) \hat{f}^{\rm RE}_{1,j}\right],\\
        \Phi^{(p),\rm hot}_{\rm diff}\left( p^{\rm hot}_{N_p+1/2}, \xi^{\rm hot}_j\right) =\\
        \left. D^{\rm hot}_{N_p+1/2,j}\frac{\partial f^{\rm hot}}{\partial p}\right|_{N_p+1/2,j} =\\
        D^{\rm hot}_{N_p+1/2,j}\frac{f^{\rm hot}_j - \hat{f}^{\rm RE}_{1,j}}
        {p_{1/2}^{\rm RE} - p^{\rm hot}_{N_p+1/2}},
    \end{gathered}
\end{equation}
with the averaged runaway distribution function
\begin{equation}
    \hat{f}^{\rm RE}_{1,j} = f^{\rm RE}_{J_j} +
    \left( \xi_j^{\rm hot} - \xi^{\rm RE}_{J_j-1} \right)
    \frac{f^{\rm RE}_{J_j} - f^{\rm RE}_{J_j-1}}
    {\xi^{\rm RE}_{J_j} - \xi^{\rm RE}_{J_j-1}},
\end{equation}
where $J_j$ denotes the smallest integer such that
$\xi^{\rm RE}_{J_j} \geq \xi^{\rm hot}_j$. The interpolation coefficients
$\delta^{(1)}_j$ should be chosen as to minimize the risk of spurious
oscillations on the grid boundary, and are therefore determined using an upwind
scheme
\begin{equation}
    \delta^{(1)}_j = \begin{cases}
        1,\qquad &\text{if }F^{\rm hot}_{N_p+1/2,j}\leq 0,\\
        0,\qquad &\text{otherwise}.
    \end{cases}
\end{equation}
The fluxes on the runaway grid are determined by combining~\eqref{eq:localflux}
and~\eqref{eq:kk:advdiffflux}, yielding
\begin{equation}
    \Phi^{(p),\rm RE}_{1/2,J} = \frac{-\sum_j \Phi^{(p),\rm hot}_{N_p+1/2,j}\Vp^{\rm hot}_{N_p+1/2,j}\overline{\Delta\xi}_{jJ}}{\Vp^{\rm RE}_{1/2,J}\Delta\xi^{\rm RE}_J}.
\end{equation}
Note that the advection and diffusion coefficients $F^{\rm hot}_{N_p+1/2,j}$ and
$D^{\rm hot}_{N_p+1/2,j}$ appear in the expressions for the fluxes on both the
hot and runaway electron grids. To conserve particles it is necessary to use
exactly the same coefficients for both grids.

%% file: tests.tex
A number of tests have been implemented for \DREAM\ in order to verify the
correctness of both the individual modules in the code, and the overall
physics modelled. For the latter, benchmarks against results in the published
literature have been performed and in this section we present the results of
three such benchmarks. The first two tests verify that the plasma conductivity
and Dreicer runaway rates are accurately computed by comparing \DREAM\
simulations to simulations with the 2D Fokker--Planck solver
\CODE~\cite{Landreman2014,Stahl2016}, and primarily validate the Fokker--Planck
collision operator used. \CODE\ only simulates homogeneous plasmas, but uses the
same test-particle collision operator as \DREAM, albeit with a finite difference
discretization in momentum and a Legendre polynomial decomposition in pitch.
The third test is to reproduce the tokamak
disruption simulations in~\cite{Vallhagen2020}, which were carried out with the
1D fluid code \GO~\cite{Smith2006,Feher2011,Papp2013}.

\subsection{Conductivity}
A typical method for validating Fokker--Planck collision operators is to solve
the Spitzer problem
\begin{equation}\label{eq:spitzer}
        eE_\parallel
    \frac{\partial f}{\partial p_\parallel} =
    C\left\{ f \right\},
\end{equation}
arising in the presence of a parallel electric field $E_\parallel$,
where $C$ is the collision operator, including collisions
between electrons and electrons, as well as electrons and ions. In the weak
electric field limit ($\Epar\ll\Ec$, with $\Ec$ the critical electric field for
runaway~\cite{Connor1975}) the current density $j$ carried by the distribution
function $f$ will be proportional to $\Epar$, with constant of proportionality
$\sigma$, i.e.\ the conductivity of the plasma at the given temperature and
effective charge. By solving equation~\eqref{eq:spitzer} for $f$, and by
extension the current density $j$, we can obtain the plasma conductivity from
the relation $\sigma = j/\Epar$.

\begin{figure}
    \centering
    \includegraphics[width=\columnwidth]{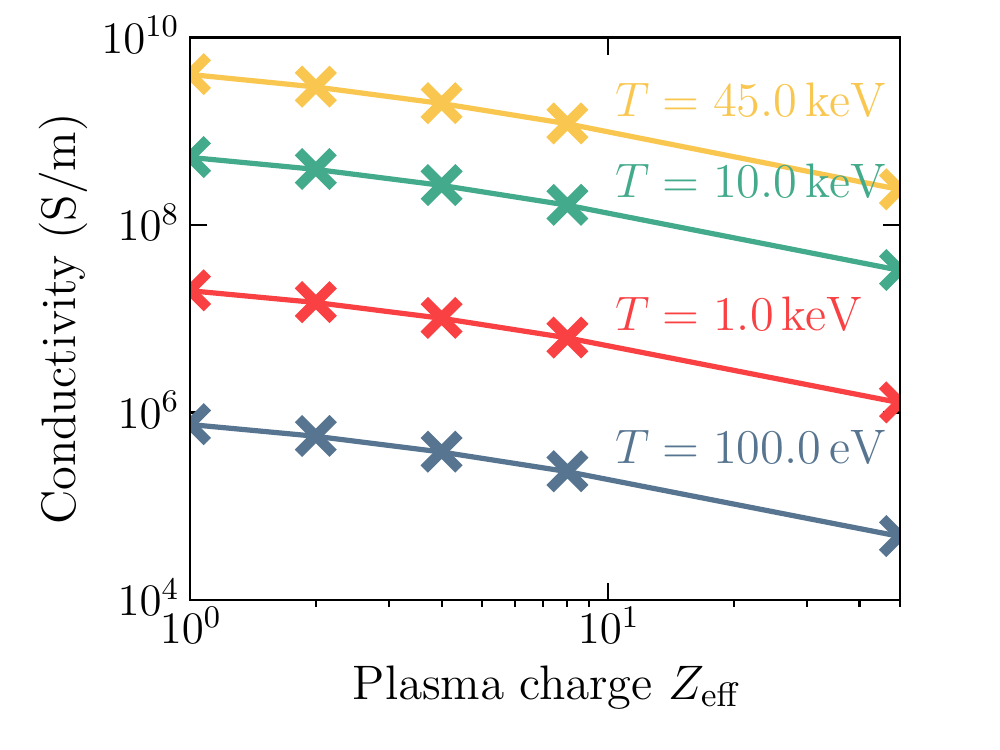}
    \caption{
        Comparison of plasma conductivity as calculated with \CODE\ (solid
        lines) and with \DREAM\ (crosses) at a few different temperatures $T$
        and effective plasma charges $Z$. All calculated conductivities match
        to within $0.3\%$.
    }
    \label{fig:conductivity}
\end{figure}

In figure~\ref{fig:conductivity}, the conductivity has been calculated with
\DREAM\ (crosses) in the {\em fully kinetic} mode at a few different
temperatures and plasma charges, and is compared to the conductivity as
calculated with \CODE\ (solid lines). Both codes implement the fully
relativistic test-particle collision operator of Ref.~\cite{Pike2016}. The
values calculated with \DREAM\ are within less than $0.3\%$ of those calculated
with \CODE, indicating that the collision operator is correctly implemented.

\subsection{Runaway rate}
\begin{figure}
    \centering
    \includegraphics[width=\columnwidth]{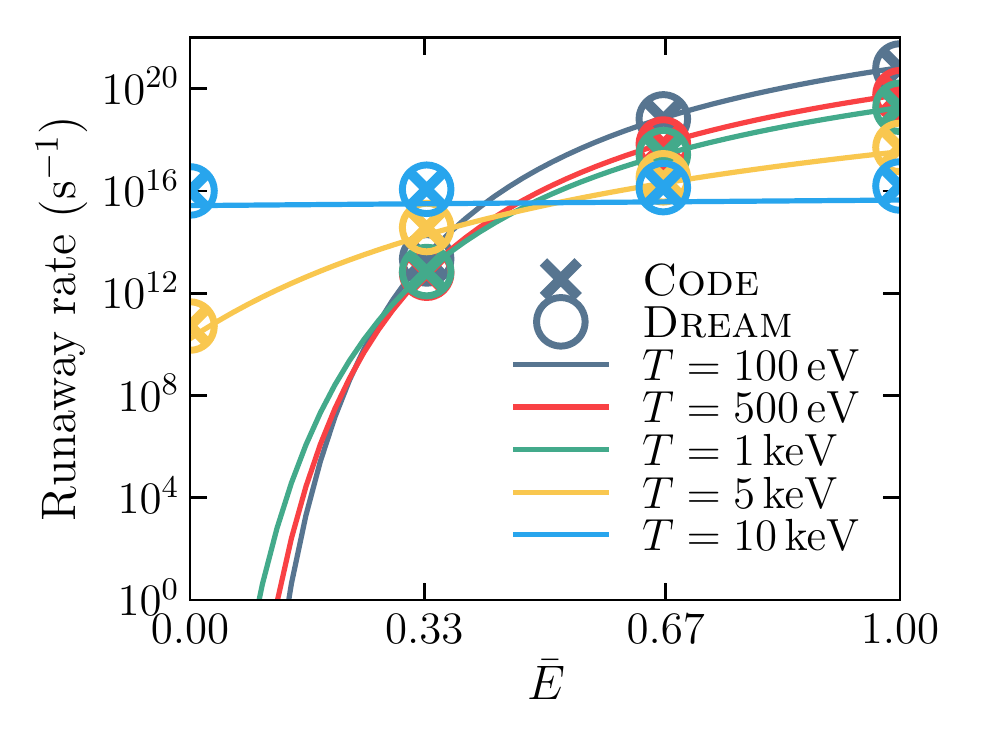}
    \caption{
        Comparison of the runaway rate as calculated with \CODE\ (circles), with
        \DREAM\ (crosses) and using the formula given in~\cite{Connor1975}
        (solid lines). The dimensionless variable
        $\bar{E} = (E-2\Ec)/(0.04\ED-2\Ec)$ is introduced so that
        $\bar{E}\in[0,1]$ covers the region of $E$ in which the runaway
        generation is mild to modest, and the linearized collision operator is
        valid. The \CODE\ and \DREAM\ runaway rates match to within $3.5\%$.
    }
    \label{fig:runawayrate}
\end{figure}

Another quantity of importance to the physics studied in \DREAM\ is the
so-called \emph{Dreicer runaway electron generation rate} obtained in a
$Z_{\rm eff}=1$ plasma when increasing $\Epar$ in the test above to
$\Epar > \Ec$~\cite{Connor1975}. In the {\em fully kinetic} mode, the number of
runaway electrons $\nRE$ is then defined as the number of particles with
momentum $p\geq\pRE$, where the runaway boundary is chosen as
$\pRE=20\sqrt{2T/mc^2}$ with $T$ the electron temperature,
and the runaway generation rate is taken as $\gamma=\partial\nRE/\partial t$.
Figure~\ref{fig:runawayrate} shows the primary runaway rate as calculated with
\DREAM\ (crosses) and \CODE\ (circles). The solid lines are calculated with the
formula given in~\cite{Connor1975}. The $x$ axis ranges from $\Epar=2\Ec$ to
$\Epar=0.04\ED$---where $\ED$ denotes the Dreicer electric
field~\cite{Dreicer1959} at which all electrons run away---corresponding to
marginal and strong runaway electron generation respectively. The \DREAM\ and
\CODE\ runaway rates match to within $3\%$.

\subsection{GO ITER simulations}\label{sec:go:iter}
To validate the coupled physics of \DREAM\, we will now present a comparison
between \DREAM\ and the simulations of ITER-like disruptions conducted
in~\cite{Vallhagen2020}. In ref.~\cite{Vallhagen2020} the effect of injecting
impurities in the plasma on the maximum runaway current in a standard ITER
scenario was studied using the 1D fluid code
\GO~\cite{Smith2006,Feher2011,Papp2013}. The \GO\ code can be considered a
predecessor of the fluid mode in \DREAM\ and uses similar models for the
background plasma evolution. Using the {\em fluid} model, as described in
section~\ref{sec:theory:re}, with a cylindrical radial grid in \DREAM, the
two codes should simulate approximately\footnote{Note, that the models for the
runaway rate and $E_c^{\rm eff}$ used in \GO\ have been replaced with
generalized versions in \DREAM, according to the expressions given
in~\ref{app:refluid}.} the same physics. While benchmarking the two codes it was discovered that some of the simulations in~\cite{Vallhagen2020} were slightly under-resolved with respect to time. While it does not lead to any qualitative differences, for this comparison we have re-run the \GO\ simulations with improved time resolution.

\begin{figure}
    \centering
    \includegraphics[width=\columnwidth]{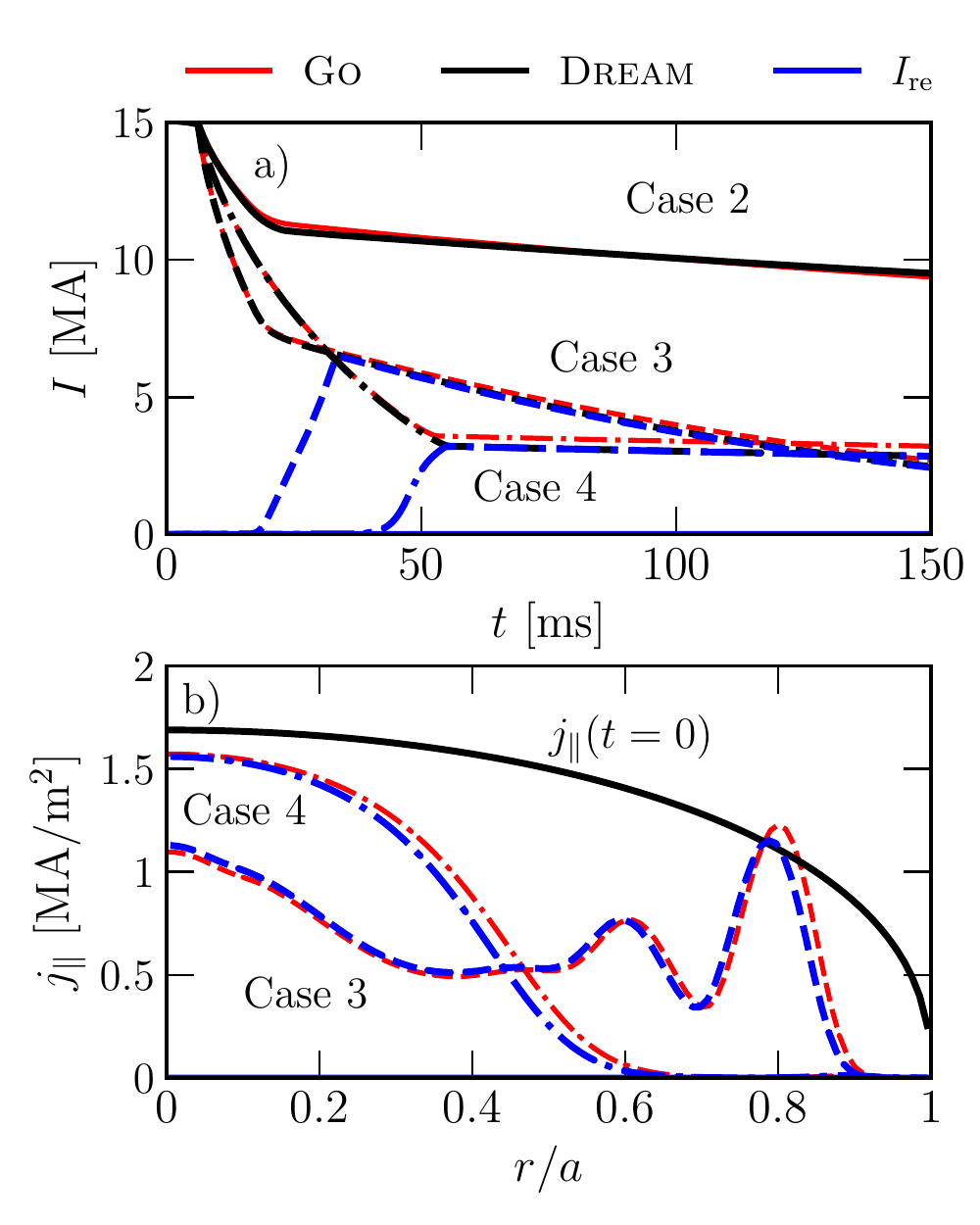}
    \caption{
        Comparison of (a) the time evolution of the total plasma current, and
        (b) maximum runaway current density profiles obtained with \DREAM\
        (black/blue) and \GO\ (red). In (a), the black curves show total plasma
        current, while blue curves show the runaway current component. In (b),
        the black curve indicates the initial current density profile.}
    \label{fig:gofig4}
\end{figure}

Figure~\ref{fig:gofig4} shows a comparison between the plasma currents and
current densities obtained for cases 2, 3 and 4 in~\cite{Vallhagen2020}. All
three cases have the same temperature, main ion density and pre-disruption
current density profiles, and only differ in the composition of the injected
material. In all cases a mixture of neutral deuterium and neon is injected and
is added instantaneously to the plasma, distributed uniformly across radii. The
amount of material injected in the different cases is summarized in
table~\ref{tab:material}.

\begin{table}
    \centering
    \caption{Amount of material injected in the ITER simulations described in
    section~\ref{sec:go:iter}.}
    \label{tab:material}
    \begin{tabular}{c|c|c}
        \textbf{Case} & \textbf{Deuterium ($10^{20}\,\si{\per\metre\cubed}$)} & \textbf{Neon ($10^{20}\,\si{\per\metre\cubed}$)}\\\hline
        Case 2 & 3 & $0.03$ \\
        Case 3 & 40 & $0.08$ \\
        Case 4 & 7 & $0.08$
    \end{tabular}
\end{table}

In all three cases considered, \DREAM\ closely reproduces both the total and
runaway plasma currents in both the thermal and current quench phases of the
disruptions, as well as the maximum runaway current density. The small
deviations between the simulation results are explained primarily by the use of
somewhat improved models for the critical electric field $E_c^{\rm eff}$ in
\DREAM.

%% file: examples.tex
To demonstrate some of the main features of \DREAM\ we will now examine two
separate disruptions in a toroidal plasma with parameters representative of
ASDEX Upgrade~\cite{Pautasso2016,Meyer2019,Pautasso2020}. In section~\ref{sec:example:baseline} we first
describe the general parameters used for all simulations and briefly recall the
differences between the electron models of \DREAM.
Sections~\ref{sec:example:fullconv} and~\ref{sec:example:slowCQ} then discuss
the results and performance of each of the electron models.

\subsection{Baseline simulation setup}\label{sec:example:baseline}
\begin{figure*}
    \centering
    \includegraphics[width=\textwidth]{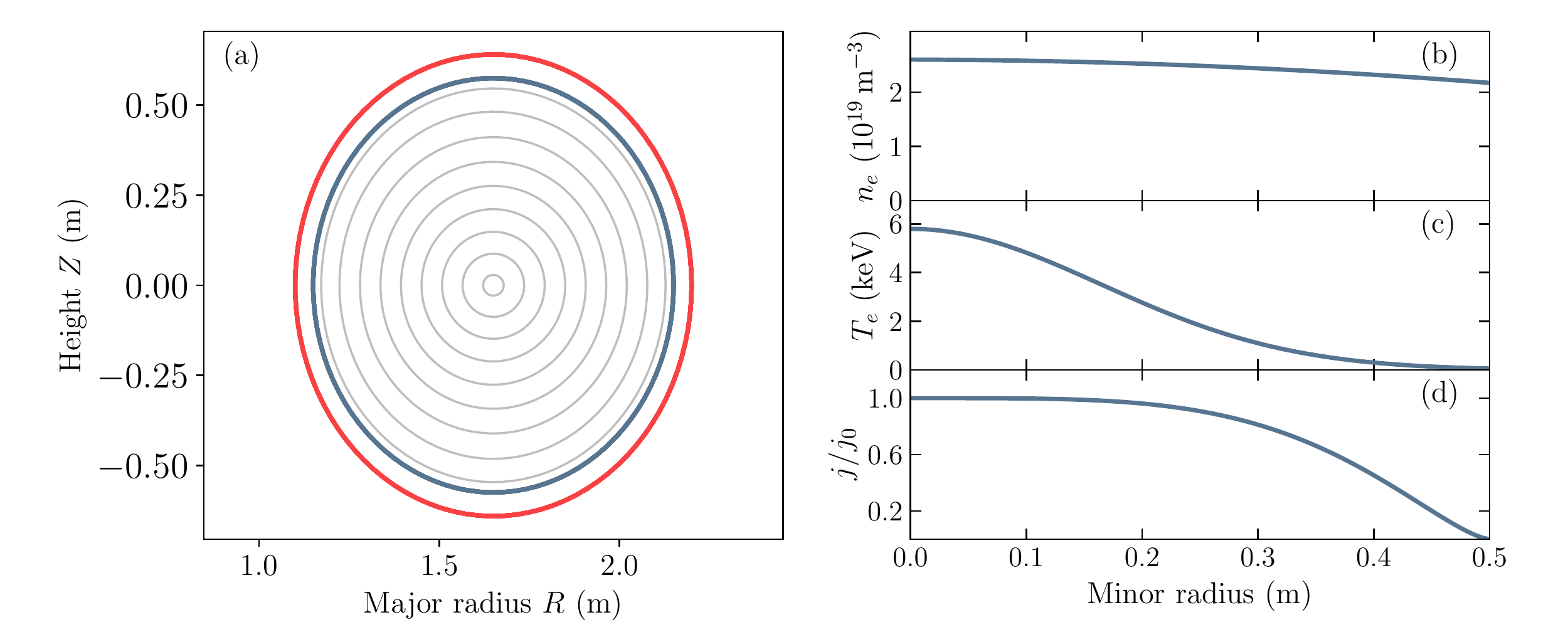}
    \caption{
        Parameters for the baseline scenario, which is used in all the
        simulations of section~\ref{sec:examples}. (a) Magnetic field flux
        surfaces (gray), with the plasma boundary shown in blue and vessel
        wall in red. (b) Initial electron density. (c) Initial electron temperature.
        (d) Initial plasma current density profile, normalized to the central
        current density $j_0=\SI{1.52}{\mega\ampere\per\metre\squared}$.
    }
    \label{fig:example:baseline}
\end{figure*}

All simulations of this section are conducted in an elongated ASDEX
Upgrade-like plasma with magnetic field and vessel parameters as shown in
table~\ref{tab:tokamak}. Figure~\ref{fig:example:baseline}a shows the
corresponding flux surfaces along with the plasma boundary (black) and
conducting vessel wall structure (red). The flux surfaces are slightly elongated
with a linearly varying elongation profile $\kappa(r) = 1 + 0.15r/a$. In
figure~\ref{fig:example:baseline}b-d radial profiles of the initial electron
density, temperature and current density are shown. The electron density is
nearly uniform, close to $n_{{\rm e},0} = \SI{2.6e19}{\per\metre\cubed}$, while
the electron temperature is peaked at
$T_{{\rm e},0}=\SI{5.8}{\kilo\electronvolt}$ on the magnetic axis and decreases
towards $\SI{60}{eV}$ near the edge. The plasma current density is
$j(r) = j_0[1-(r/a)^4]^{3/2}$, with $j_0=\SI{1.52}{MA\per\metre\squared}$ chosen
to give the desired total initial plasma current $I_{{\rm p},0} = \SI{800}{kA}$.

\begin{table}
    \centering
    \caption{
        Magnetic field and vessel parameters used in the simulations of
        section~\ref{sec:examples}.
    }
    \label{tab:tokamak}
    \begin{tabular}{r|l}
        \textbf{Parameter} & \textbf{Value}\\\hline
        Major radius $\Rm$ & $\SI{1.65}{\metre}$\\
        Minor radius $a$ & $\SI{0.5}{\metre}$\\
        Wall radius $b$ & $\SI{0.55}{\metre}$\\
        Elongation at edge $\kappa(a)$ & $1.15$\\
        Toroidal magnetic field $B_0$ & $\SI{2.5}{\tesla}$\\
        Initial plasma current $I_{{\rm p},0}$ & $\SI{800}{\kilo\ampere}$
    \end{tabular}
\end{table}

In the following sections we will insert a combination of neutral deuterium and
neutral argon into the plasma outlined above. The material is assumed to be
instantly distributed uniformly across the plasma. Using four different electron
models, we follow the evolution of the plasma using as it cools down due to
radiation losses as well as a prescribed diffusive heat transport,
according to~\eqref{eq:heat}, with $D_W=\SI{4000}{m^2\per s}$ in
section~\ref{sec:example:fullconv} and $D_W=\SI{1000}{m^2\per s}$ in
section~\ref{sec:example:slowCQ}. The four models
used are the \emph{fully kinetic}, the \emph{superthermal} and the
\emph{isotropic} models described in section~\ref{sec:theory:hot}, as well as a
\emph{fluid} model similar to the one used in the \GO\ 
code~\cite{Vallhagen2020,Feher2011, elongation}, which we briefly commented on
in section~\ref{sec:theory:re}. The main differences between these models can be
briefly summarized as follows: in the {\em fully kinetic} model, both cold and
hot electrons are modelled kinetically while runaway electrons are modelled as a
fluid; in the {\em superthermal} model, only hot electrons are modelled
kinetically, while cold and runaway electrons are modelled as fluids; the 
{\em isotropic} model makes the same assumptions as the superthermal model, but
uses an angle-averaged kinetic equation and evolves only the energy distribution
of the hot electrons; in the {\em fluid} model, only thermal bulk and runaway
electron populations are followed, both as fluids.

Another important difference separating the {\em fully kinetic}/{\em fluid} and
{\em superthermal}/{\em isotropic} models is the treatment of the cold electron
temperature, \Tcold, which is used both in the test-particle collision
operator and in the ion rate equations~\eqref{eq:ionrateequation}. In all models
the temperature is evolved according to equation~\eqref{eq:heat}, and the
electron distribution is initialized at equilibrium with the temperature of the 
hot initial plasma. In the {\em fully kinetic} and {\em fluid} models---which do
not distinguish between cold and warm electrons---the temperature \Tcold\ starts
at the initial warm plasma temperature, and rapidly falls as cold impurities are
inserted in the plasma. In contrast, in the {\em superthermal} and
{\em isotropic} models, \Tcold{} starts at almost zero, corresponding to the 
temperature of the injected impurities. The injected electrons will promptly
form a Maxwellian at a significantly lower temperature 
than that of the pre-disruption plasma, and the hot electrons will predominantly 
slow down in free-free collisions with this cold Maxwellian. Because of these 
differences, the {\em fully kinetic} and {\em fluid} models can be expected to
agree relatively well, while the {\em superthermal} and {\em isotropic} models
could be expected to differ from the former two in some cases.

The cooling-down process occurring as impurities are injected into the hot plasma
leads to an inherently non-linear evolution for the
distribution function as it transitions from containing two Maxwellian electron
populations---the injected electrons at a very low temperature and the initial
bulk electrons at a warm, but gradually cooling, temperature---into a state
dominated by a single Maxwellian electron population. The {\em superthermal} and
{\em isotropic} models provide a numerically efficient way of capturing some of
this dynamic without resorting to a fully non-linear collision operator.

\begin{table}
    \centering
    \caption{
        Typical wall clock times for the simulations conducted in
        section~\ref{sec:examples} on an Intel Xeon desktop computer with
        a total of $2\,800$ time steps.
    }
    \begin{tabular}{c|c|c}
        \textbf{Model} & $N_r\times N_\xi\times N_p$ & \textbf{Wall time}\\
        \hline
                Fluid & $15\times 1\times 1$ & $\SI{25}{seconds}$\\
                Isotropic & $15\times 1\times 80$ & $\SI{1}{minute}$ $\SI{40}{seconds}$\\
                Superthermal & $15\times 68\times 80$ & $\SI{1}{hour}$ $\SI{7}{minutes}$\\
                Fully kinetic & $15\times68\times 140$ & $\SI{4}{hours}$
                    \end{tabular}
\end{table}

\subsection{Full-conversion scenario}\label{sec:example:fullconv}
\begin{figure}
    \centering
    \includegraphics[width=\columnwidth]{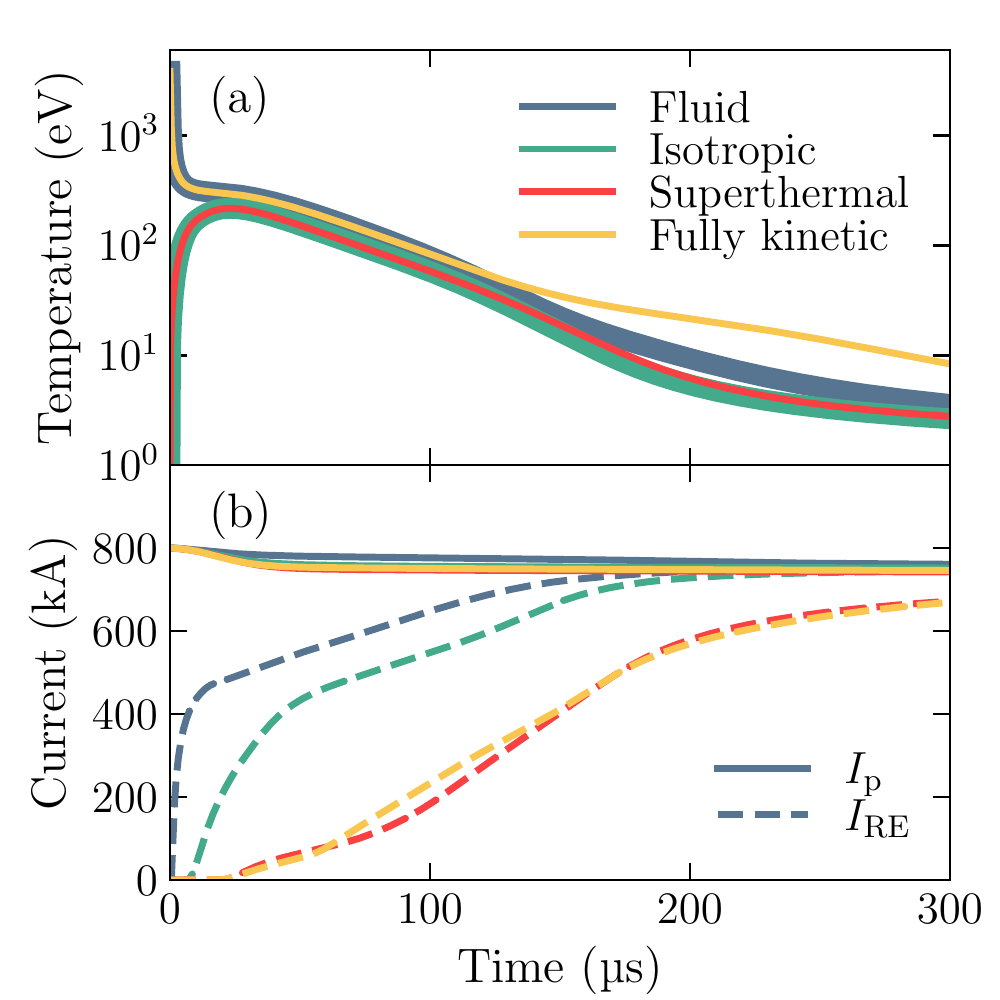}
    \caption{
        Time evolution of (a) cold electron temperature at $r/a=0.3$, (b) plasma
        current (solid) and runaway current (dashed) in the full-conversion
        scenario.
    }
    \label{fig:scenario1}
\end{figure}

In this scenario we initiate the plasma as described in
section~\ref{sec:example:baseline} and insert neutral deuterium and argon with
radially uniform densities $n_{\rm D}=n_{\rm Ar}=\SI{2.6e19}{\per\metre\cubed}$
at $t=0$. As shown in figure~\ref{fig:scenario1}, the resulting
temperature and current dynamics are very fast, with much of the thermal quench
completing in about one hundred microseconds. Due to the rapid cooling, a
significant fraction of electrons remain hot at the onset of the current quench, 
leading to a seamless conversion of ohmic current into superthermal and then
runaway current via the hot-tail mechanism. By the end of the current quench, almost all of the original
current has been converted into runaway current in all three models.

Some differences are observed in the evolution of the temperature and total
runaway current in the {\em fully kinetic}/{\em fluid} and
{\em superthermal}/{\em isotropic} models, although the total plasma current
reached is almost exactly the same in all cases. Note that the temperature shown
for the {\em superthermal} and {\em isotropic} models is that of the cold
injected electrons, while the temperature shown for the {\em fully kinetic} and
{\em fluid} models is that of the (initially warm) bulk electrons. The main
reason for the differences in temperature evolution is the electron-ion heat
exchange $Q_{ij}$ in equation~\eqref{eq:collheat}, which explicitly depends on
the relative temperature difference between electrons and ions. Since the
initial value of $\Tcold$ differs in the four models, so does the heat
transferred from the ions to the electrons, and hence also the detailed
evolution of $\Tcold$. The fact that the runaway currents still agree well is
due to the thermal quench being rapid in all models, which leads to the
formation of a significant hot electron population that is eventually
accelerated and runs away.

\begin{figure}
    \centering
    \includegraphics[width=\columnwidth]{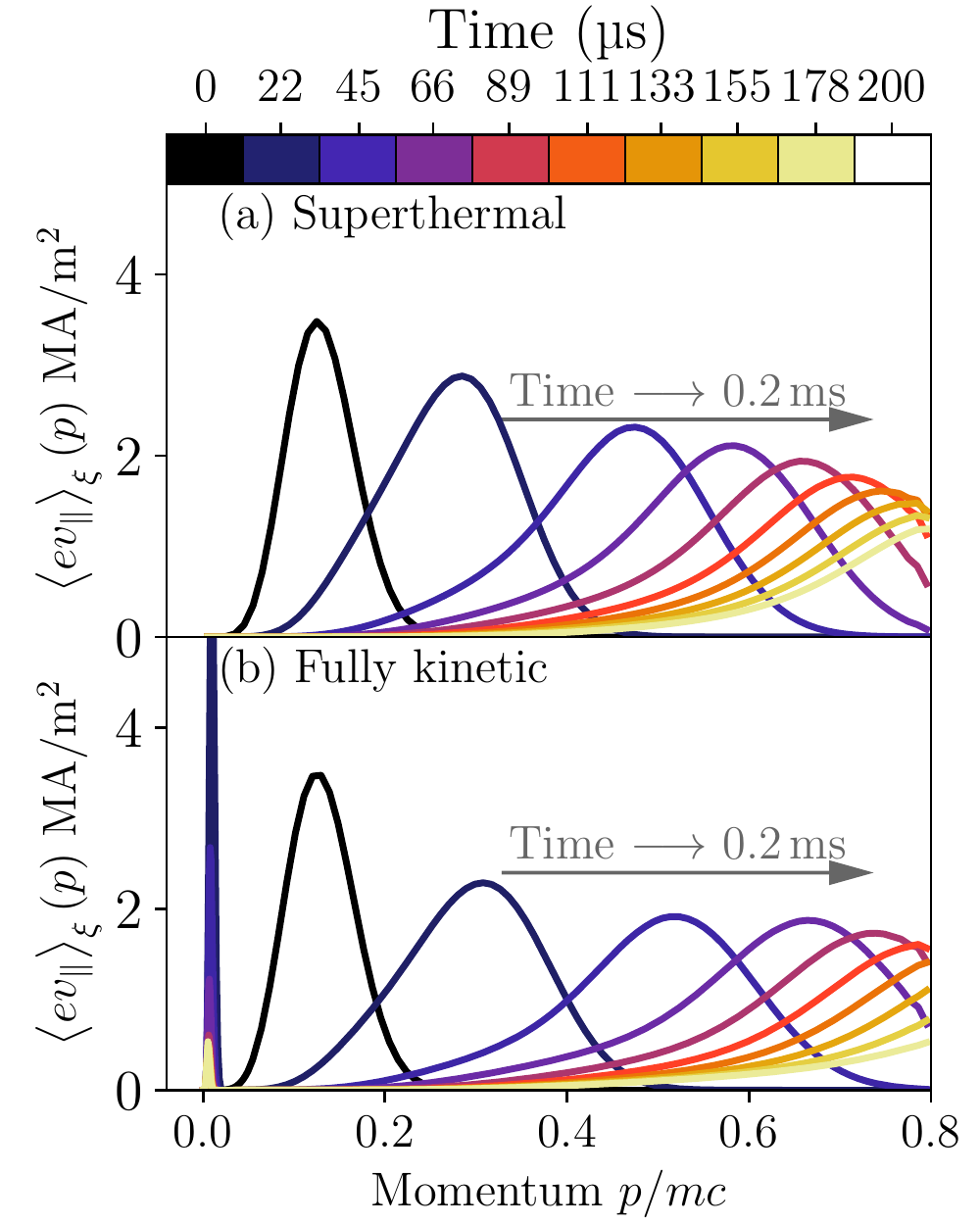}
    \caption{
        Time evolution of the $\xi$-averaged contribution to the parallel
        current from (a) the hot electron distribution function
        $f_{\rm hot}(r,p,\xi)$ in the {\em superthermal} model, and (b) from the
        electron distribution function $f(r,p,\xi)$ in the {\em fully kinetic}
        model. The latter contains both the ohmic (seen as a large peak near
        $p=0$) and the hot electron current.
    }
    \label{fig:scenario1:currmom}
\end{figure}

Figure~\ref{fig:scenario1:currmom} illustrates the evolution of the current
carried by the distribution functions in the {\em superthermal} and
{\em fully kinetic} models. In the former, the ohmic current is obtained from
Ohm's law (as in equation~\eqref{eq:jomega}) and the distribution function only
carries superthermal current, while in the latter the distribution function also
contains the thermal bulk electrons, and thus also carries ohmic current (seen
as a sharp peak near $p=0$ in figure~\ref{fig:scenario1:currmom}b). During the
current quench, the remaining superthermal electrons---which carry a significant
fraction of the total current---are accelerated to even higher momenta and run
away. The absence of the thermal bulk in the {\em superthermal} model permits
the momentum grid resolution to be nearly halved, making the {\em superthermal}
model computationally efficient while accurate in capturing the hot-tail
formation.

\subsection{Slow disruption scenario}\label{sec:example:slowCQ}
\begin{figure}
    \centering
    \includegraphics[width=\columnwidth]{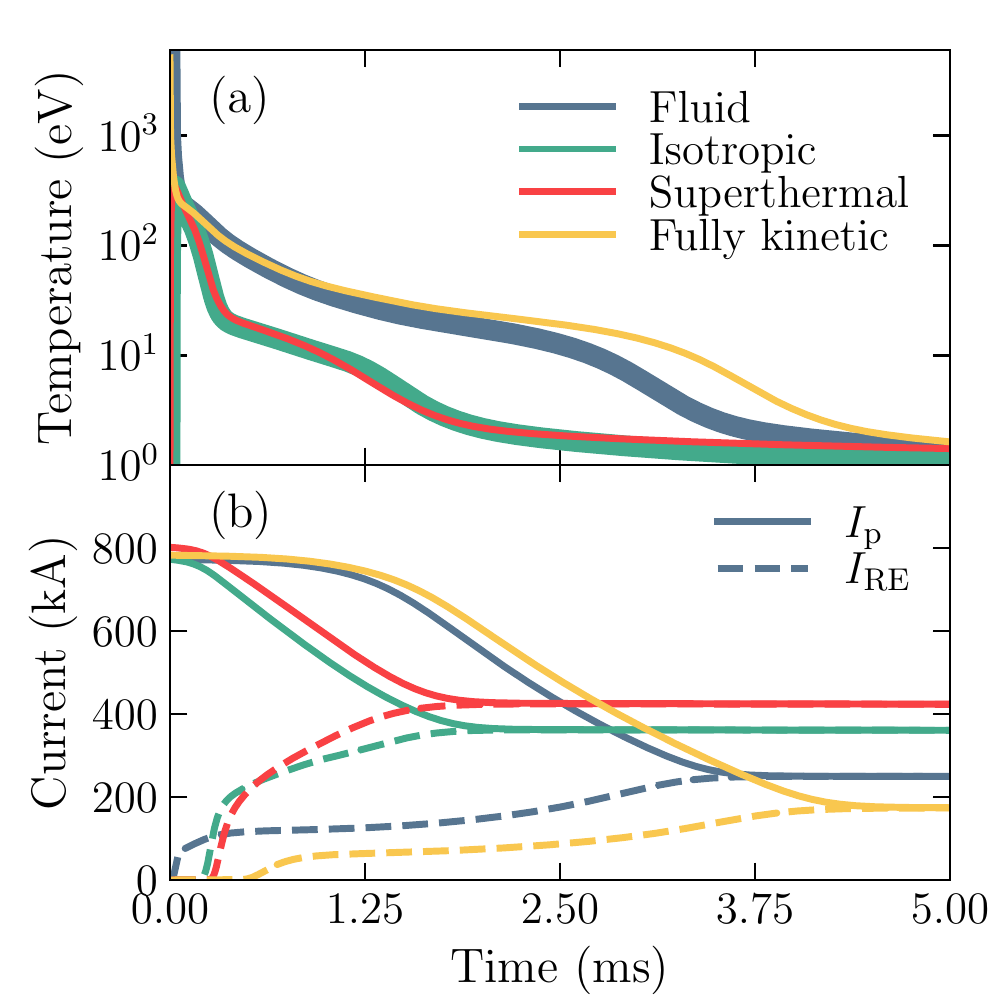}
    \caption{
        Time evolution of (a) cold electron temperature at $r/a=0.17$, (b)
        plasma current (solid) and runaway current (dashed) in the slow
        disruption scenario of section~\ref{sec:example:slowCQ}.
    }
    \label{fig:slowdisr}
\end{figure}

In this scenario we initiate the plasma as described in
section~\ref{sec:example:baseline} and insert neutral deuterium and argon with
radially uniform densities $n_{\rm D}=\SI{5.2e20}{\per\metre\cubed}$ and
$n_{\rm Ar} = \SI{5.2e18}{\per\metre\cubed}$ respectively at $t=0$. The
resulting disruption occurs over a relatively long time, with the current quench
lasting for up to $\SI{4}{ms}$, depending on the model used, as shown in
figure~\ref{fig:slowdisr}b. As in section~\ref{sec:example:fullconv}, the
temperature shown in figure~\ref{fig:slowdisr}a for the {\em superthermal} and
{\em isotropic} models is that of the cold injected electrons, while the temperature
shown for the {\em fully kinetic} and {\em fluid} models is that of the
(initially warm) bulk electrons. In contrast to the scenario of
section~\ref{sec:example:fullconv}, this scenario reveals significant
differences between the different models used.  The {\em fluid} and
{\em fully kinetic} models mostly agree with each other, as do the
{\em isotropic} and {\em superthermal} models, but when comparing the
{\em superthermal} and {\em fully kinetic} models---the two most advanced
models---the final runaway currents are found to deviate by a factor of two.

The deviations between the {\em fluid}/{\em fully kinetic} and
{\em isotropic}/{\em superthermal} models in figure~\ref{fig:slowdisr} are
consequences of the self-consistent plasma evolution, although the origin of the
different evolutions can be traced to the disparate definitions of the
temperature used. Several terms and coefficients depend explicitly on the
temperature $\Tcold$, including the ion rate coefficients in
equation~\eqref{eq:ionrateequation}, the collisional energy transfer
term~\eqref{eq:collheat}, and the radial heat diffusion term, and as such,
differing dynamics are to be expected in the brief initial phase of the TQ when
$\Tcold$ differs significantly between the two groups of models. However, it is
only if one or more of these temperature-dependent terms are dominant during the
early TQ phase that the final runaway current should be significantly impacted.
In the scenario of figure~\ref{fig:slowdisr} it turns out that the radial heat
diffusion term plays an important role in the {\em fluid} and
{\em fully kinetic} models early during the TQ, causing the $\Tcold$ profile to
be flattened. This in turn alters the behaviour of the electric field, which
typically grows rapidly in response to the decreased conductivity when the
temperature drops.

\begin{figure}
    \centering
    \includegraphics[width=\columnwidth]{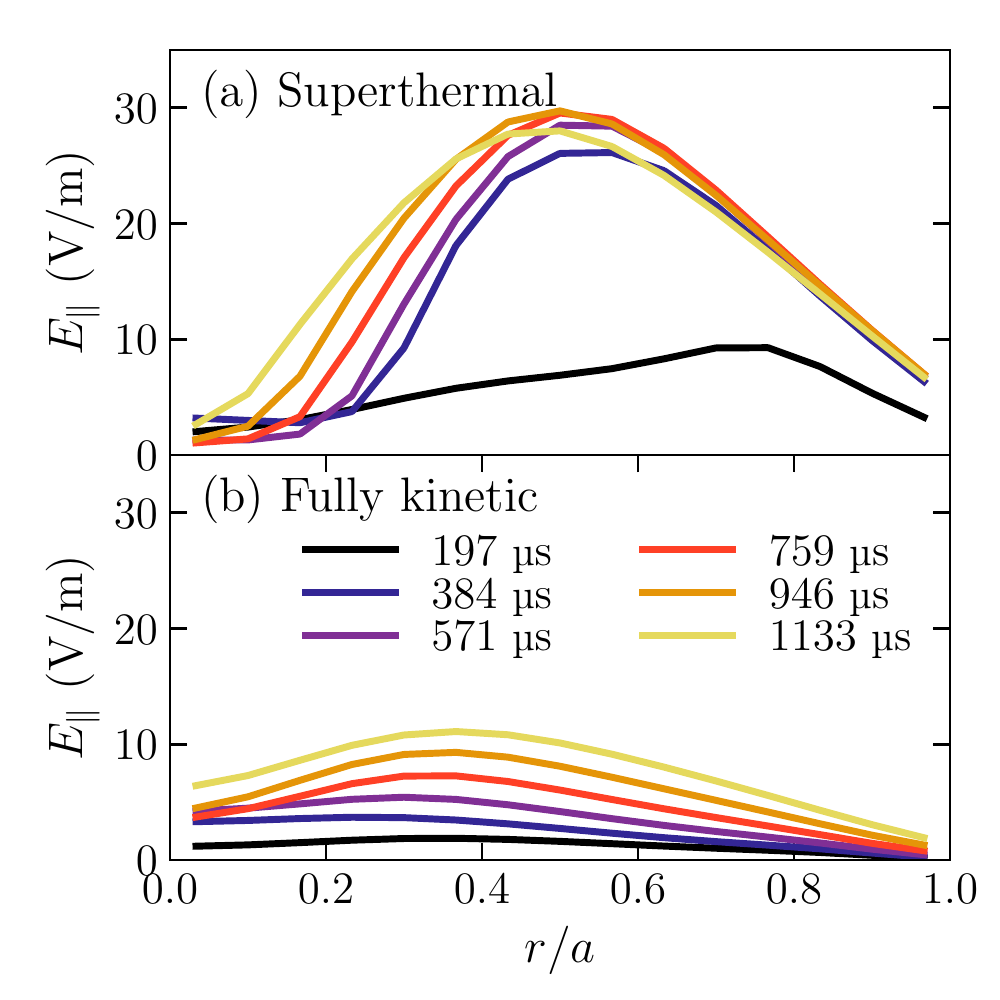}
    \caption{
        Electric field as a function of radius at a few times during the TQ
        phase in the scenario of section~\ref{sec:example:slowCQ} using (a) the
        {\em superthermal} model, and (b) the {\em fully kinetic} model. The
        large difference in initial temperature between the models results in a
        significantly slower electric field evolution in the {\em fully kinetic}
        model.
    }
    \label{fig:slowdisrE}
\end{figure}

With the {\em fluid} and {\em fully kinetic} models, the relatively strong
radial heat diffusion causes the temperature to decrease rapidly in the centre
of the plasma, but also to be slightly raised at outer radii. As a result, the
remaining thermal energy is radiated away more slowly, allowing the ohmic
current to be maintained for a longer time, and thus delaying the increase of
the electric field. Figure~\ref{fig:slowdisrE} shows the electric field
evolution in the {\em superthermal} and {\em fully kinetic} models during this
early phase of the disruption. The strong electric fields during the early phase
of the disruption leads to a significant conversion of hot electrons to
runaways. In the {\em fully kinetic} case, figure~\ref{fig:slowdisrE}b, the
slower electric field evolution does not allow for as many hot electrons to be
immediately converted into runaways, but partially compensates for this later on
during the disruption by driving more production of runaways through the
avalanche mechanism. The increased avalanche generation in the
{\em fully kinetic} model is however not sufficient to fully compensate for the
early hot-tail generation in the {\em superthermal} model.

The deviations between the {\em fluid} and {\em fully kinetic} models, as well
as the {\em isotropic} and {\em superthermal} models, stem almost entirely from
the differences in how the hot-tail generation is modelled. Both the {\em fluid}
and {\em isotropic} models utilise approximations to the Fokker--Planck
treatment of the hot-tail mechanism used in the {\em superthermal} and
{\em fully kinetic} modes. As a result, the number of runaway electrons
generated via the hot-tail mechanism is slightly over- and underestimated,
respectively, in the {\em fluid} and {\em isotropic} modes.

Finally, a comment on the physics fidelity of the four considered models  is due. The
{\em fluid} and {\em isotropic} models are direct approximations of the more advanced models and are less reliable, although the
results here suggest that their results for the temperature and current evolution are reasonably close to the more
advanced kinetic models. As for the {\em superthermal} and {\em fully kinetic} models,
it is difficult to clearly state that one is more reliable than the other. The {\em fully kinetic} model uses a linearized collision operator and in the early phase of the thermal quench, the process may be inherently non-linear. When a large amount of
impurities are injected, the electron distribution will briefly be constituted
by two Maxwellians at different temperatures, which the {\em fully kinetic}
model is not equipped to handle. The {\em superthermal} model, on the other
hand, is derived with this exact situation in mind and therefore provides a
better approximation of the processes. The {\em superthermal} model
is however still an approximation to the full disruption physics, and its
assumptions, e.g.\ that a large number of impurities are present, are not
necessarily always well satisfied. To verify the hot-tail models considered
here, a relativistic non-linear collision operator, as used in
Refs.~\cite{norse,Daniel2020}, coupled to a self-consistently evolving
background plasma, is therefore needed. This will be considered in future work.

%% file: appGeometry.tex
In an axisymmetric tokamak of major radius $\Rm$, with ions uniformly distributed on the flux surfaces, the poloidal dependence of the coefficients of the transport equation~\eqref{eq:transport} is naturally described in terms of three independent quantities $\eta_1 = B/B_{\rm min}$, $\eta_2 = R/\Rm$, $\eta_3 = |\nabla r|^2$, as well as the pitch-angle dependent variable $\eta_4 = \xi/\xi_0 = \sqrt{[1-(1-\xi_0^2)B/B_{\rm min}]/\xi_0^2}$. These have been introduced such that they all equal unity in a circular plasma, characterized by $\Rm=\infty$, which simplifies solutions in this limit. The minimum magnetic field on a flux surface has been denoted $B_{\rm min}(r) = \min(\sqrt{\boldsymbol{B}(r,\,\theta)^2}\,|\, \theta)$, with $r$ a flux surface label.

Flux surface integrals are efficiently carried out via 
\begin{equation}
\int_{-\pi}^\pi X \, \dd \theta \approx \sum_{i=1}^N w_i X(\theta_i),
\end{equation}
evaluated with a Gauss-Legendre quadrature rule where the four independent angle-dependent quantities $\BA{\eta_n}_{n=1}^4$ as well as the Jacobian $\mathcal{J} = 1/|\nabla \varphi \cdot (\nabla r \times \nabla \theta)|$ are precomputed at the nodes $\theta_i$, where $N = 10$ is found to typically be sufficient for relative errors $<10^{-4}$. For bounce integrals over trapped orbits, the integration limits $\theta_{b1}(\xi_0)$ and $\theta_{b2}(\xi_0)$ depend on pitch and the metric has an integrable singularity ($\sim 1/\sqrt{\theta_{b1,b2}^2 - \theta^2}$) at the boundary, and a Chebyshev-Gauss quadrature is employed instead. In this case, the precomputed values must be separately evaluated on individual poloidal angle meshes for each trapped orbit, in contrast to passing orbits for which the same poloidal angle mesh can be reused.

The bounce-orbit metric \Vp{} contains a logarithmic singularity on the trapped-passing boundary $\xi_0 = \pm \xi_T$. In order to resolve these singular points, instead of using the midpoint rule to estimate the cell average in the finite-volume methods, for cells containing a singular point we instead carry out the pitch average
\begin{align}
\oint X(\xi_i, \,\theta) \, \dd \theta\mapsto \frac{1}{\Delta \xi_i} \int_{\xi_{i-1/2}}^{\xi_{i+1/2}} \dd \xi \oint \dd \theta \, X(\xi,\,\theta)
\end{align}
as a double integral using the adaptive QAWS routine of QUADPACK, which is designed for integrals with endpoint singularities.  For brevity, we have not written out other arguments on which the integrand may depend.

\DREAM{} supports the use of an analytic up-down symmetric geometry where the flux surfaces are parametrized according to
\begin{align}
\boldsymbol{x} &= R\hat{R} + z\hat{z}, \nonumber \\
R &= \Rm + \Delta(r) + r\cos[\theta + \delta(r)\sin\theta], \nonumber \\
z &= r \kappa(r) \sin\theta, \nonumber \\
\hat{R} &= \cos\varphi \hat{x} + \sin\varphi\hat{y}, \nonumber \\
\boldsymbol{B} &= G(r) \nabla \varphi + \frac{\psi_\mathrm{ref}'(r)}{2\pi} \nabla \varphi \times \nabla r.
\end{align}
Here, the Shafranov shift $\Delta$, elongation $\kappa$ and triangularity $\delta$ parametrize the shape of the flux surfaces, and $G$ and the reference poloidal flux gradient $\psi_\mathrm{ref}'$ (that is left independent of the poloidal flux $\psi$, evolved dynamically in the equation system) determine the strength of the toroidal and poloidal components of the magnetic field, respectively. In this geometry, the Jacobian $\mathcal{J}$ and scale factor $|\nabla r|^2$ are given by
\begin{align}
\hspace{-3mm}\mathcal{J} &=  \kappa r R \biggl\{ \cos(\delta\sin\theta) + \Delta'\cos\theta + \sin\theta\sin[\theta+\delta\sin\theta]\nonumber \\
&  \times \left[\frac{r\kappa'}{\kappa} + \delta\cos\theta\left(1+\frac{r\kappa'}{\kappa}-\frac{r\delta'}{\delta} \right) \right] \biggr\}, \nonumber \\
\hspace{-3mm} |\nabla r|^2 &=\frac{\kappa^2 r^2R^2}{\mathcal{J}^2}\Bigl[  \frac{1}{\kappa^2}(1+\delta\cos\theta)^2\sin^2(\theta+\delta\sin\theta) \nonumber \\
&+ \cos^2\theta\Bigr].
\end{align}

%% file: appKinEq.tex
\subsection{Heat transport due to fast-electron transport}\label{app:kineq:DW}
The heat transport associated with the electron particle transport is obtained by integrating the diffusive transport term of the kinetic equation with the diffusion coefficient~\eqref{eq:rechester-rosenbluth} over a Maxwell-J\"uttner distribution function,
\begin{align}
f_M &= \frac{\ncold}{4\pi \Theta K_2(1/\Theta) } e^{-\gamma/\Theta}, \nonumber \\
\frac{\partial f_M}{\partial r} &= \frac{f_M}{\Tcold\Theta}\left( \gamma-3\Theta - \frac{K_1(1/\Theta)}{K_2(1/\Theta)} \right) \frac{\partial \Tcold}{\partial r},\\
\Theta &= \frac{\Tcold}{m_e c^2},
\end{align}
where we do not keep the contribution from $\partial \ncold/\partial r$ since electron density cannot be transported independently of the ions; the electron density profile is set by quasineutrality. As such, we assume that the heat transport acts to flatten the temperature profile. Therefore, the heat diffusion coefficient is given by the energy moment ($m_e c^2(\gamma-1)$)  
\begin{align}
D_W &= \frac{2}{3\ncold}\int_0^\infty \mathrm{d}p\,\frac{f_M}{\Theta^2}(\gamma-1)\nonumber\\ &\times \left(\gamma - 3\Theta - \frac{K_1(1/\Theta)}{K_2(1/\Theta)}\right) 
\int_{-1}^1 \mathrm{d}\xi_0 \, \frac{\Vp}{V'} \BA{D^{rr}}.
\end{align}
For the Rechester-Rosenbluth model~\eqref{eq:rechester-rosenbluth} the pitch integral is given by 
\begin{equation}
\int_{-1}^1 \mathrm{d}\xi_0 \, \frac{\Vp}{V'} \BA{D^{rr}} = 2\pi p^2 v\FSA{ \frac{B}{B_\mathrm{max}} }\pi q \Rm c \left(\frac{\delta B}{B}\right)^2,
\end{equation}
yielding
\begin{align}
D_W &= \frac{2}{3}\FSA{\frac{B}{B_\mathrm{max}}} \pi q \Rm c\left(\frac{\delta B}{B}\right)^2  \frac{1}{e^{1/\Theta}K_2(1/\Theta)} \nonumber \\
&\times\left[  (2+3\Theta)\left(1-\frac{K_1}{K_2}\right) + 3\Theta^2 \right] \\
&\approx \frac{2\pi}{3} q \Rm c\left(\frac{\delta B}{B}\right)^2\sqrt{\frac{18\Theta}{\pi}}\left[1 - \frac{5}{8}\Theta \right]\FSA{\frac{B}{B_\mathrm{max}}} , \nonumber
\end{align}
with the last line corresponding to the non-relativistic limit $\Theta \ll 1$, retaining the leading-order relativistic correction.

\subsection{Reduced kinetic equation}\label{app:kineq:reduced}
For the {\em isotropic} electron mode, described in section~\ref{sec:theory:hot},
a reduced form of the usual kinetic equation detailed in
section~\ref{sec:theory:kinetic} is used, analogous to
Ref.~\cite{RosenbluthPutvinski1997}. To derive the reduced equation, we first
introduce the ordering parameter $\delta$ and assume $\nu_D\sim\delta^0$
and $E\sim\delta^1$, with all other terms being of order $\delta^2$. Writing
$f=f_0+\delta f_1 + \mathcal{O}(\delta^2)$ and grouping the kinetic equation by
order in $\delta$, we obtain a set of new equations governing the evolution of
the distribution function:
\begin{subequations}
    \begin{equation}\label{eq:app:reduced0}
        \delta^0:\quad
        \frac{\partial}{\partial\xi_0}\left[
            \Vp\BA{\frac{\xi^2}{\xi_0^2}\frac{\Bmin}{B}}
            \left(1-\xi_0^2\right)\frac{\partial f_0}{\partial\xi_0}
        \right] = 0
    \end{equation}
    \begin{equation}\label{eq:app:reduced1}
        \begin{gathered}
            \delta^1:\quad
            e\BA{\Epar\xi}\left[
                \frac{1}{p^2}\frac{\partial}{\partial p}\left(p^2f_0\right) +
                \frac{1}{p\xi_0}\frac{\partial}{\partial\xi_0}\left[
                    \left(1-\xi_0^2\right)f_0
                \right]
            \right] =\\
            = \frac{\nu_D}{2\Vp}\frac{\partial}{\partial\xi_0}\left[
                \Vp\BA{\frac{\xi^2}{\xi_0^2}\frac{\Bmin}{B}}
                \left(1-\xi_0^2\right)\frac{\partial f_1}{\partial\xi_0}
            \right]
        \end{gathered}
    \end{equation}
    \begin{equation}\label{eq:app:reduced2}
        \begin{gathered}
            \delta^2:\quad
            	\frac{\partial f_0}{\partial t} = \frac{1}{\Vp}\frac{\partial}{\partial p}\biggl[
            		\Vp \biggl( -\BA{\hat{A}^p}f_0 \\
            			- e\BA{\Epar\xi}f_1 + \BA{D^{pp}}\frac{\partial f_0}{\partial p}
				\biggr)
			\biggr] + \frac{1}{\Vp}\frac{\partial \Vp F^{\xi_0}}{\partial\xi_0} \\
            + \frac{1}{\Vp}\frac{\partial}{\partial r}\left[
                \Vp\left(
                    -\BA{A^r} f_0 + \BA{D^{rr}}\frac{\partial f_0}{\partial r}
                \right)
            \right] + \BA{S},
        \end{gathered}
    \end{equation}
\end{subequations}
where in the last equation, $\hat{A}^p$ denotes the momentum advection that is not due 
to electric-field acceleration, and $F^{\xi_0}$ denotes the net pitch flux, which will not affect
the final result.
Solving the $\delta^0$ equation~\eqref{eq:app:reduced0} first yields the leading
order solution
\begin{equation}
    f_0 = f_0\left(t,r,p\right),
\end{equation}
i.e.\ $f_0$ is isotropic. Substituting this into the
next order equation~\eqref{eq:app:reduced1} and integrating the result over
$\xi_0$ from $-1$ to $\xi_0$ yields
\begin{equation}
    \begin{aligned}
        -\pi h\left(\xi_0\right) p^2 &V'\frac{e\EdotB}
        {\nu_D\Bmin}
        \frac{\partial f_0}{\partial p} =\\
        &\Vp\BA{\frac{\xi^2}{\xi_0^2}\frac{\Bmin}{B}}
        \left(1-\xi_0^2\right)\frac{\partial f_1}{\partial\xi_0},
    \end{aligned}
\end{equation}
with the function
\begin{equation}
    \begin{aligned}
        h\left(\xi_0\right) &= 2\int_{\xi_0}^1\dd\xi_0\,\xi_0 \mathcal{H}(\xi_0) \\
        &= \begin{cases}
            1-\xi_T^2,&\text{trapped},\\
            1-\xi_0^2,&\text{passing},
        \end{cases}
    \end{aligned}
\end{equation}
where the step function $\mathcal{H}$ was defined in~\eqref{eq:H func}.
A solution for $f_1$ can then be obtained by isolating
$\partial f_1/\partial\xi_0$ and integrating over $\xi_0$ once more,
giving
\begin{equation}\label{eq:app:kineq:f1}
    \begin{gathered}
    f_1\left(t,r,p,\xi_0\right) =
        -\pi p^2\frac{eV'\EdotB}{\nu_D\Bmin}
        \frac{\partial f_0}{\partial p}\times\\
        \times\int_{-1}^{\xi_0} \frac{h\left(\xi_0'\right)\,\dd\xi_0'}
        {\Vp\BA{\xi^2\Bmin^2/(\xi_0^2B^2)}\left(1-\xi_0'^2\right)}.
    \end{gathered}
\end{equation}

With $f_0$ and $f_1$ determined, we may now use the $\delta^2$
equation~\eqref{eq:app:reduced2} to obtain the final reduced kinetic equation.
We first multiply both sides of equation~\eqref{eq:app:reduced2} by $\Vp$ and
integrate over all $\xi_0$. Because of this, the pitch angle scattering term, as
well as the $\xi_0$ component of the electric field term, vanish due to the
factors of $1-\xi_0^2$ which are zero at $\xi_0=\pm1$. The slowing down and
transient terms remain mostly unaffected by the integration, only picking up a
factor of
\begin{equation}
    \begin{gathered}
        \int_{-1}^{1}\dd\xi_0\,\Vp =
        2\pi\int_0^{2\pi}\dd\phi\int_{-1}^1\dd\xi_0\oint\dd\theta\,\sqrt{g} =\\
        = 2\pi\int_0^{2\pi}\dd\phi\int_0^{2\pi}\dd\theta\left(
            \int_{-1}^{-\xi_T} + \int_{\xi_T}^1
        \right) \dd\xi_0\,\sqrt{g} =\\
        = 2\pi p^2\int_0^{2\pi}\dd\phi\int_0^{2\pi}\dd\theta\,\mathcal{J}\times\\
            \times\left(\int_{-1}^{-\xi_T}+\int_{\xi_T}^1 \right)
            \frac{B}{\Bmin}\frac{\xi_0}{\xi}\,\dd\xi_0
        = 4\pi p^2 V',
    \end{gathered}
\end{equation}
while the electric field term becomes
\begin{equation}\label{eq:app:kineq:Eterm}
    \begin{gathered}
        2\pi \frac{V'e\EdotB}{\Bmin}
        \frac{\partial}{\partial p}
        \int_{-1}^1 \dd\xi_0\,p^2\xi_0 f_1 \mathcal{H}\left(\xi_0\right) =\\
        = 
        -\pi \frac{V'\left(e\EdotB\right)^2}{\Bmin^2}
        \frac{\partial}{\partial p}
        \int_{-1}^1\dd\xi_0\,\frac{p^2\xi_0}{\nu_D} \mathcal{H}\left(\xi_0\right)\times\\
        \times\int_{-1}^{\xi_0}\frac{-h\left(\xi_0'\right)\dd\xi_0'}
        {\FSA{\xi/\xi_0'}\left(1-\xi_0'^2\right)}
    \end{gathered}
\end{equation}
where the expression~\eqref{eq:app:kineq:f1} for $f_1$ was substituted.
After interchanging the order of integration, the $\xi_0'$ integral ranges over
the passing region only due to the step function, while the $\xi_0$ integral is
recognized as half the function $h(\xi_0')$. Since we will then only evaluate
$h(\xi_0')$ in the passing region, we may replace it by its actual value there,
$h(\xi_0'\in\text{passing}) = 1-\xi_0'^2$. Furthermore, since the integrand is
even in $\xi_0'$ in the passing region, we may replace the integral with two
times the integral ranging from $\xi_T$ to $1$:
\begin{equation}
    \begin{gathered}
        \eqref{eq:app:kineq:Eterm} =
            -\pi \frac{V'\left(e\EdotB\right)^2}{\Bmin^2}
            \int_{\xi_T}^1\frac{1-\xi_0'^2}{\FSA{\xi/\xi_0'}}\,\dd\xi_0'\times\\
            \times\frac{\partial}{\partial p}\left(
                \frac{p^2}{\nu_D}\frac{\partial f_0}{\partial p}
            \right).
    \end{gathered}
\end{equation}
The electric field term, which was purely advective in the original kinetic
equation, has now become purely diffusive after the averaging procedure.

For the radial transport terms the $\xi_0$ integral will only apply to the
advection and diffusion coefficients, and so we can make use of the relation
\begin{equation}
    \begin{aligned}
        \int_{-1}^1\dd\xi_0\,\Vp\BA{X} &=
        \int_{-1}^1\dd\xi_0\,\Vp\frac{2\pi}{\Vp}\int_0^{2\pi}\dd\phi\oint\dd\theta\sqrt{g}X =\\
        &= 2\pi\int_0^{2\pi}\dd\phi\int_{-1}^1\dd\xi_0\oint\sqrt{g}X =\\
        &= 2\pi p^2\int_0^{2\pi}\dd\phi\int_0^{2\pi}\dd\theta\int_{-1}^1\dd\xi\,X =\\
        &= 4\pi p^2\FSA{\frac{1}{2}\int_{-1}^1\dd\xi\,X }
        \equiv 4\pi p^2\FSA{X}_{\xi},
    \end{aligned}
\end{equation}
where in the next-to-last step we used that $\Bmin\xi\,\dd\xi=B\xi_0\,\dd\xi_0$.
In what remains, we will use the notation $\FSA{X}_{\xi}$ to denote the
combined flux surface and pitch average of a quantity $X$. 

After the steps above, we obtain the final reduced kinetic equation by dividing
all terms by $\int\Vp\dd\xi_0 = 4\pi p^2V'$, yielding
\begin{equation} \label{eq:app:transport}
    \begin{gathered}
        \frac{\partial f_0}{\partial t}  = \frac{1}{p^2}\frac{\partial}{\partial p}\biggl[p^2\biggl(
        -\FSA{\hat{A}^p}_\xi f_0 + \left(\FSA{D^{pp}}_\xi + \mathcal{D}_E\right)\frac{\partial f_0}{\partial p} \biggr)\biggr] \\
		+ \frac{1}{V'}\frac{\partial}{\partial r}\left[
            V'\left( 
                -\FSA{A^r}_{\xi}f_0 +
                \FSA{D^{rr}}_{\xi}\frac{\partial f_0}{\partial r}
            \right)
        \right]
        + \FSA{S}_{\xi}, \\
        \mathcal{D}_E = f_{\rm p}\frac{(e\EdotB)^2}{3\FSA{B^2}\nu_D(p)}.
    \end{gathered}
\end{equation} 
Here we have introduced $\mathcal{D}_E$ as the momentum diffusion coefficient
representing the electric-field acceleration, and
the effective passing fraction, denoted $f_{\rm p}$, is defined as
\begin{equation}
    f_{\rm p} = \frac{3}{4}\FSA{\frac{B^2}{\Bmin^2}}
    \int_{\xi_T}^1\frac{1-\xi_0'^2}{\FSA{\xi/\xi_0'}}\,\dd\xi_0'.
\end{equation}

\paragraph{Current density}
The current density corresponding to the distribution evolved by the reduced
kinetic equation can be calculated from equation~\eqref{eq:jhot}. Since
$f_0$ is isotropic, $f_1$ is the lowest-order component of $f$ to carry any
current, and the current density therefore becomes
\begin{equation}
    \begin{gathered}
        \frac{j_{\rm iso}}{B} = \pi e^2\frac{\EdotB}{\Bmin^2}
        \int\dd p\dd\xi_0\,\frac{vp^2}{\nu_D}\mathcal{H}\left(\xi_0\right)\xi_0
        \frac{\partial f_0}{\partial p}\times\\
        \times\int_{-1}^{\xi_0} \frac{h\left(\xi_0'\right)\dd\xi_0'}
        {\FSA{\xi/\xi_0'}\left(1-\xi_0^2\right)}.
    \end{gathered}
\end{equation}
Just as for the electric field term, we interchange the order of integration in
$\xi_0$ and $\xi_0'$ and recognize that the $\xi_0$ integral yields a factor of
$(1/2)h(\xi_0')$. The current density is hence given by
\begin{equation}\label{eq:app:kineq:j}
    \frac{j_{\rm iso}}{B} = \frac{2\pi e^2f_{\rm p}}{3}\frac{\EdotB}{\FSA{B^2}}
    \int\dd p\frac{vp^2}{\nu_D}\frac{\partial f_0}{\partial p}.
\end{equation}
An undesirable property of~\eqref{eq:app:kineq:j} is that it allows for the
current density to grow larger than $ecn_e$, which is the current density
expected when all electrons travel at the speed of light along magnetic field
lines. Exceeding this value may destabilize the solver, and so we adjust
for this behaviour by smoothly matching the limit of all electrons travelling
parallel to magnetic field lines, in which case the current density is given by
\begin{equation}
    \frac{j_{\rm iso,2}}{B} = 4\pi e\,\mathrm{sgn}(\EdotB)\int\dd p\,vp^2f,
\end{equation}
where $\mathrm{sgn}(x)$ denotes the sign function. The actual current density
used in the simulation is then taken as the matched formula
\begin{equation}
    j = \frac{j_{\rm iso} j_{\rm iso,2}}{ \hypot{j_{\rm iso}}{j_{\rm iso,2}}},
\end{equation}
corresponding approximately to the smallest of the two approximations.

%% file: appRunawayFluid.tex
A widely used formula for the avalanche growth rate in a pure plasma was derived
by Rosenbluth \& Putvinski~\cite{RosenbluthPutvinski1997}, accounting for
geometric effects in a large-aspect-ratio tokamak. This formula has later been
generalized to partially ionized 
plasmas~\cite{MartinSolis2015highZ, hesslowNF2019}, which has also been applied
to runaway generation due to Compton scattering and tritium beta
decay~\cite{MartinSolis2017,Vallhagen2020}. Here, we present generalized fluid
formulae for the runaway generation rate that extends the validity of previous
work to axisymmetric tokamak geometry with shaped surfaces of arbitrary aspect
ratio.

\subsection{Analytic runaway rate from source function $S$}
Runaway production mechanisms other than Dreicer and hot-tail generation---such
as large-angle collisions, beta decay of tritium and Compton scattering---are
typically modelled by a source function defined in the particle phase 
space $S=S(t,\,\bb{x},\,\bb{p})$. A general procedure for deriving the corresponding 
fluid runaway rate from such a source function follows from the ordering made
in~\ref{app:kineq:reduced} in the superthermal limit, where  $D^{pp}$ is negligible.
We also assume runaway formation to occur in 
quasi-steady state and that radial transport occurs on longer time scales,
by ordering $\partial/\partial t\sim A^r\sim D^{rr}\sim\delta^3$. Doing so 
casts the second-order equation~\eqref{eq:app:reduced2} into the form
\begin{equation}\label{eq:app:anaS}
    \frac{1}{p^2}\frac{\partial}{\partial p}\left[
        p^2\left( \FSA{A^p}_\xi f + \mathcal{D}_E\frac{\partial f}{\partial p} \right)
    \right] + \FSA{S}_\xi = 0,
\end{equation}
In the steady-state fluid picture, since $\langle A^p\rangle_\xi$ and
$\mathcal{D}_E$ become constant in the limit $v\to c$, the rate at which new
runaway electrons are generated can be defined as the particle flux to
$p=\infty$,
\begin{equation}
    \frac{\partial \FSA\nRE}{\partial t} = -4\pi p^2\left(
        \FSA{A^p}_\xi f + \mathcal{D}_E\frac{\partial f}{\partial p}
    \right)_{p=\infty}.
\end{equation}
Note that the purpose of the source function $S$ in~\eqref{eq:app:anaS} is to
create new free electrons at any energy, not necessarily runaway electrons. For a
newly created electron to become a runaway, it must be successfully accelerated
into the runaway region, and the probability for this is determined by the
advective and diffusive processes dominating the electron dynamics. By
integrating equation~\eqref{eq:app:anaS} from $p$ to $\infty$ we obtain a
relation between the runaway generation rate, the source term $S$ and the
advection-diffusion processes determining the probability for runaway:
\begin{equation}\label{eq:app:advdiffS}
    \begin{gathered}
        p^2\left( \FSA{A^p}_\xi f + \mathcal{D}_E \frac{\partial f}{\partial p} \right) =\\
        -\frac{1}{4\pi}\frac{\partial \FSA\nRE}{\partial t} +
        \int_p^\infty \dd p'\,p'^2 \FSA{S(p')}_\xi.
    \end{gathered}
\end{equation}
To solve this equation for the runaway generation rate, we introduce the
integrating factor
\begin{equation}\label{eq:Gfunc}
    G = -\int_p^\infty\frac{\FSA{A^p}_\xi(p')}{\mathcal{D}_E(p')}\,\dd p',
\end{equation}
and divide equation~\eqref{eq:app:advdiffS} by $p^2 D^{pp}$ to obtain
\begin{equation}\label{eq:app:intfac}
    \frac{\partial\ee^G f}{\partial p} =
    \frac{\ee^G}{p^2D_E}\left(
        -\frac{1}{4\pi}\frac{\partial \FSA\nRE}{\partial t} +
        \int_p^\infty\dd p'\,p'^2\FSA{S}_\xi\left(p'\right)
    \right).
\end{equation}
If we assume that $f$ is well-behaved, so that
$\ee^Gf|_{p=\infty} = \ee^Gf|_{p=0}=0$, where the latter follows from the fact
that $\lim_{p\to 0} G = -\infty$, we can integrate~\eqref{eq:app:intfac} over
all momenta:
\begin{equation}
    \begin{gathered}
        \frac{1}{4\pi}\frac{\partial \FSA\nRE}{\partial t}
        \int_0^\infty\dd p\frac{\ee^G}{p^2\mathcal{D}_E}
        =\\
        =
        \int_0^\infty\dd p\frac{\ee^G}{p^2\mathcal{D}_E}
        \int_p^\infty\dd p'\,p'^2\FSA{S}_\xi\left(p'\right)
        =\\
        =
        \int_0^\infty\dd p\,p^2 \FSA{S}_\xi\left(p\right)
        \int_0^p\dd p'\,\frac{\ee^{G\left(p'\right)}}{p'^2\mathcal{D}_E\left(p'\right)}
    \end{gathered}
\end{equation}
Solving for $\partial \FSA\nRE/\partial t$ then yields
\begin{equation}
	\begin{gathered} \label{eq:source rate}
    		\frac{\partial\FSA\nRE}{\partial t} =
    		4\pi\int_0^\infty\dd p\,p^2\FSA{S}_\xi\left(p\right)h\left(p\right) \\
    		= \FSA{\int h(p)S(t,\bb{x},\bb{p})\, \dd\bb{p}},
    \end{gathered}
\end{equation}
where
\begin{equation}
    h\left(p\right) =
    \frac{\int_0^p\dd p'\frac{\ee^G}{p'^2\mathcal{D}_E}}
    {\int_0^\infty\dd p'\frac{\ee^G}{p'^2\mathcal{D}_E}}
\end{equation}
can be interpreted as the probability for an electron created with momentum $p$
to run away. In the non-relativistic limit, for a fully ionized plasma with constant Coulomb
logarithm, $\FSA{A^p}_\xi$ is proportional to $1/p^2$, allowing an exact integration 
of the runaway probability
\begin{equation}
h(p) = e^G = \exp\left[ 
	-\frac{3}{4}\frac{\FSA{B^2}p^2\nu_s\nu_D}{f_{\rm p}(e\EdotB)^2}
\right]
\end{equation}
Since the exponent varies rapidly with momentum, $p^2\nu_s\nu_D \propto 1/p^4$, 
the runaway probability $h$ is well approximated by a step function at the critical 
momentum $p_\star$ defined by
\begin{equation}
	\begin{gathered}
		p_\star^2\nu_s(p_\star)\nu_D(p_\star) = f_{\rm p}\frac{(e\EdotB)^2}{\FSA{B^2}}.
	\end{gathered}
\end{equation}
The validity of the runaway rates can be extended to the near-threshold regime $E\sim E_c$ and to weak pitch scattering $\nu_D \sim 0$ by defining a matched formula for the critical runaway momentum $p_c$ according to~\cite{hesslowNF2019}
\begin{equation}
p_c = \left(\frac{\bar\nu_s(p_\star)\bar\nu_D(p_\star) + 4\bar\nu_s(p_\star)^2}{f_{\rm p}e^2({\EdotB}/{\sqrt{\FSA{B^2}}} - \Eceff)^2}\right)^{1/4},
\end{equation}
where $\bar\nu_s = p^3\nu_s/\gamma^2$ and $\bar \nu_D = p^3\nu_D/\gamma$ denote normalized collision frequencies that depend only weakly on momentum.
In this expression \Eceff\ denotes the effective critical field, described in~\ref{sec:Eceff}, and $p_c = \infty$ when the electric field is smaller than \Eceff. Compared to ref.~\cite{hesslowNF2019}, the numerator includes a factor of $\bar\nu_s^2$ which increases the accuracy of the formula for weakly ionized low-$Z$ plasmas.
In \DREAM, the runaway rate due to source functions are evaluated using equation~\eqref{eq:source rate} with the runaway probability
\begin{equation}
h(p) = H(p-p_c),
\end{equation}
where $H$ denotes the Heaviside step function.
The source function $S$ due to Compton and tritium decay are modelled as in ref.~\cite{Vallhagen2020}.

\subsection{Evaluation of critical electric field}\label{sec:Eceff}
The critical electric field for runaway generation is the weakest electric field
required for net acceleration of any electron in the plasma to occur. The
original expression for this field, derived by Connor and
Hastie~\cite{Connor1975}, only considered electric field acceleration and
deceleration due to collisional friction. However, it has been experimentally
observed that the actual, or effective, critical electric field is likely larger
than the value given by Connor and Hastie. Several theoretical studies have
since generalized original Connor and Hastie expression to also account for the
effect of bremsstrahlung and synchrotron radiation, as well as the partial
ionization of atoms. In \DREAM, we use the method described
in~\cite{Hesslow2018b}, further generalized to tokamak geometry. Below, we
derive the equation which must be minimized to obtain the effective critical
electric field and describe how it is implemented in a computationally
efficient manner.

\subsubsection*{Theory}
Following ref.~\cite{Hesslow2018b}, we consider the kinetic equation
\begin{equation}\label{eq:app:refluid:kineq}
    \begin{aligned}
    \frac{\partial f}{\partial t} &=
        -\frac{1}{\Vp}\frac{\partial}{\partial p}\left( \Vp \left\{ A^p \right\} f \right)\\
        &+ \frac{1}{\Vp}\frac{\partial}{\partial\xi_0}\left[
            \Vp\left(
                -\left\{A^{\xi_0}\right\} f +
                \left\{D^{\xi_0\xi_0}\right\}\frac{\partial f}{\partial\xi_0}
            \right)
        \right],
    \end{aligned}
\end{equation}
where the bounce-averaged advection-diffusion coefficients $\{A^p\}$,
$\{A^{\xi_0}\}$ and $\{D^{\xi_0\xi_0}\}$ contain the effects of electric field
acceleration, partial ionization and radiation losses (bremsstrahlung and
synchrotron) as described in~\ref{app:kineq}. Near the critical
momentum for runaway acceleration it is expected that pitch fluxes dominate
over energy fluxes, i.e.\ $\{A^{\xi_0}\}\sim\{D^{\xi_0\xi_0}\}\gg\{A^p\}$,
allowing us to obtain an expression for the steady-state pitch distribution
\begin{equation}\label{eq:app:refluid:fPitch}
    \begin{gathered}
        0 = -\left\{A^{\xi_0}\right\} f + \left\{D^{\xi_0\xi_0}\right\}\frac{\partial f}{\partial\xi_0},\\
        \implies f\left(\xi_0\right)\propto \exp\left[ -\int_{\xi_0}^1 \frac{\left\{ A^{\xi_0} \right\}}{\left\{ D^{\xi_0\xi_0} \right\}}\,\dd\xi_0' \right].
    \end{gathered}
\end{equation}
The exponent in the expression for $f$ can be written in the form
\begin{equation}\label{eq:app:refluid:Ag}
    \frac{\left\{A^{\xi_0}\right\}}{\left\{D^{\xi_0\xi_0}\right\}} = Ag\left(\xi_0\right),
\end{equation}
where in the cylindrical theory $A = 2eE/(p\nu_D)$ and $g=1-\xi_0$. In the
general case, we have
\begin{equation*}
    \begin{aligned}
        A &= \frac{2e\EdotB}{p\nu_D B_{\rm min}},\\
        g &= \begin{cases}
            H\left(\xi_0, 1\right), &\quad \xi_T < \xi_0 \leq 1\\
            H\left(\xi_T, 1\right), &\quad -\xi_T \leq \xi_0 \leq \xi_T\\
            H\left(\xi_T, 1\right) + H\left(\xi_0, -\xi_T \right), &\quad -1\leq \xi_0 < -\xi_T
        \end{cases}
    \end{aligned}
\end{equation*}
with the auxiliary function
\begin{equation}\label{eq:app:refluid:H}
    H\left(\xi_1,\xi_2\right) = \int_{\xi_1}^{\xi_2}\frac{\xi_0}{\FSA\xi}\dd\xi_0.
\end{equation}

Next, after substituting~\eqref{eq:app:refluid:fPitch} back into the kinetic
equation~\eqref{eq:app:refluid:kineq}, the kinetic equation is integrated over
$\dd\xi_0 \Vp/V'$ to yield
\begin{equation}
    \frac{\partial F_0}{\partial t} + \frac{\partial\left( U(p)F_0 \right)}{\partial p} = 0,
\end{equation}
where the distribution-weighted bounce averaged momentum flux---or net
acceleration---$U(p)$ is
\begin{equation}\label{eq:app:refluid:Up}
    U(p) = \frac{\int\Vp\left\{A^p\right\}f\,\dd\xi_0}{\int\Vp f\,\dd\xi_0}.
\end{equation}
The effective critical electric field $E_c^{\rm eff}$ is then defined as the
minimum value of the electric field for which there exists a real solution to
$U(p)=0$, that is
\begin{equation}\label{eq:app:refluid:opt}
    E_c^{\rm eff} = \min\left(\left.\frac{\EdotB}
        {\sqrt{\FSA{B^2}}} \right| U(p)=0 \right).
\end{equation}

\subsubsection*{Implementation}
The calculation of $U(p)$ typically requires repeated evaluation of three nested
integrals, two of which stand inside an exponential function, and is hence not
entirely straightforward to implement efficiently. To speed up evaluation we use
splines to evaluate the function $g(\xi_0)$ in~\eqref{eq:app:refluid:Ag} as well
as bounce averages of the advection coefficient $A^p$. For the function
$g(\xi_0)$, we construct splines representing the integrand
$\xi_0/\FSA\xi$ in~\eqref{eq:app:refluid:H} by evaluating the
integrand on a uniform $\xi_0$ reference grid, which subsequently allows the
function $H(\xi_1,\xi_2)$ to be efficiently evaluated using routines for exact
integration of splines.

The advection coefficient $A^p$ can be factorised into
$A^p = \sum_i a_i^p(p)\hat{A}_i^p(\xi_0,\theta)$ with the sum $i$ taken over 
equation terms contributing to the force balance, and where the prefactor depends only on
momentum and the remainder only on pitch and poloidal angle. The bounce average
of the pitch-dependent part of the advection coefficients, $\{\hat{A_i}^p\}$, are
then spline interpolated onto a uniform pitch grid in the interval
$\xi_0\in [0, 1]$, since all advection operators considered are either symmetric
or anti-symmetric in $\xi_0$. The distribution-weighted bounce average of the
coefficients $\hat{A}^p$ are then spline interpolated to the uniformly sampled
variable $X = A^2/(1+A)^2\in[0,1]$ (in which the functions are smoothly varying
all the way up to the limit $A=\infty$, corresponding to all runaways having
$\xi=1$), where $A$ is the inverse pitch distribution width parameter given
in~\eqref{eq:app:refluid:Ag}, allowing rapid evaluation of the acceleration
function $U$.

The root of~\eqref{eq:app:refluid:opt} is then solved for as a nested
optimization problem with two layers. In the outer layer, a solution is sought
to the one-dimensional root finding problem
\begin{equation}
    U_e\left( \EdotB \right) = 0,
\end{equation}
where $U_e$ is the maximum of $U(p)$ with respect to $p$ at a given electric
field \EdotB, i.e.\ the strongest acceleration
experienced by any particle. The problem is solved using an unbounded secant
method, assuming for the initial guess that $E_c^{\rm eff} / E_c^{\rm tot}$
is constant in time, where $E_c^{\rm tot}$ denotes the classical critical
electric field given in Ref.~\cite{Connor1975}, evaluated with $n_e$ being the
density of both free and bound electrons.

In the inner layer, the strongest acceleration at any momentum
\begin{equation}
    U_e = \min_p\left[ -U\left( p \right) \right],
\end{equation}
is determined. This problem is solved using Brent's method~\cite{Brent1971} from
the GNU Scientific Library~\cite{GSL}. To ensure fast and robust convergence of
the method, the algorithm is applied to the interval $p_{\rm opt}(1\pm0.02)$,
where $p_{\rm opt}$ is the minimum from the previous solve. If the interval does
not contain the minimum, it is expanded in steps of $20\%$ until a minimum is
enclosed. Expansion of the interval is typically needed less than once in a
thousand solves.

\subsection{Dreicer runaway rate}
Dreicer runaway generation in \DREAM\ can be modelled using the neural network presented in Ref.~\cite{hesslow2019dreicer}, which was trained on kinetic simulations in
cylindrical geometry for a wide range of ion compositions, temperatures and electric fields. Since the Dreicer rate is exponentially sensitive to the normalized electric field $E/E_D$, with $E_D = n e^3  \ln\Lambda /(4\pi\varepsilon_0^2 \Tcold)$ the Dreicer field, the runaway rate will typically be sharply peaked near the time when $E/E_D$ takes its maximum value. During the disruption, before significant runaway generation has occurred, the current is mainly ohmic so that $E \sim \eta j \propto \Tcold^{-3/2}$, whereas the Dreicer field scales like $\Tcold^{-1}$. Therefore $E/E_D \propto \Tcold^{-1/2}$, indicating that Dreicer generation occurs when the temperature approaches its minimum value, typically in the 5-10\,eV range in a disruption scenario. At such low temperatures, it has been shown that trapping effects are significantly suppressed~\cite{mcdevitt2019}, and therefore we neglect such effects by evaluating the neural network at the average parallel electric field $E_\parallel = \EdotB/\sqrt{\FSA{B^2}}$. 

\subsection{Hot-tail formation}
\DREAM\ implements a fluid description of hot-tail formation which is similar to the method described in Ref.~\cite{Smith2008}, but differs in the counting of runaways, resulting in increased accuracy at high plasma charge. Different models for hot-tail generation, including the method implemented here, have been described in detail and compared to each other in a recent report~\cite{Svenningsson2020}.

The hot-tail generation is calculated by starting with the reduced \emph{isotropic} kinetic equation given by~\eqref{eq:app:transport}, but ordering the transport and electric-field acceleration as small, $A^r \sim D^{rr} \sim \mathcal{D}_E \sim \delta^3$. Although the electric field term will be needed to evaluate the runaway rate, the initial slowing down occurs while electric fields are still weak and have little impact on the energy spectrum of hot electrons. The resulting equation in this limit is given by
\begin{equation}\label{eq:app:hottail dist}
\frac{\partial f_0}{\partial t} + \frac{1}{p^2}\frac{\partial}{\partial p}\left(p^2\FSA{\hat{A}^p}_\xi f_0\right) = 0,
\end{equation}
where collisional momentum diffusion $D^{pp}$ can be neglected due to the low temperatures during the thermal quench. The solution to this slowing-down problem was given in Ref.~\cite{Smith2008} when $\hat{A}^p$ was dominated by collisional friction in a fully ionized, non-relativistic, plasma: 
\begin{align}
f_0 &= \frac{n_0(r)}{\pi^{3/2}p_{Te0}^3}\exp\left[- \left(p^3+3\tau\right)^{2/3}/p_{Te0}^2 \right], \nonumber \\
\label{eq:app:analytic hot dist}
\tau(t) &= \int_0^t\nu_c \,\dd t,  \\
\nu_c &= 4\pi \ln\Lambda_0  r_0^2 c \ncold, \nonumber
\end{align}
where $\tau$ is the time-integrated collision frequency, and the solution was subject to the initial condition of a Maxwellian at temperature $n_0$ and temperature $T_0 = p_{Te0}^2 m_e c^2/2$ at time $t=0$.

The runaway rate is obtained by considering how the electric-field term would contribute to equation~\eqref{eq:app:hottail dist}. The ratio of advection to diffusion coefficients $\FSA{\hat{A}^p}_\xi/\mathcal{D}_E \propto p\nu_s \nu_D \sim 1/p^5$, meaning that for momenta beyond a critical value $p_0$, the diffusion term is going to be dominant and will lead to rapid runaway acceleration. This momentum is defined in terms of the vanishing net momentum flux
\begin{equation}\label{eq:app:hottail pc}
\left. \FSA{\hat{A}^p}_\xi f_0 + \mathcal{D}_E\frac{\partial f_0}{\partial p} \right|_{p=p_0} = 0,
\end{equation}
which depends on the instantaneous distribution.
If we consider the density of electrons having momentum $p>p_0$ as runaways, integration of equation~\eqref{eq:app:hottail dist} yields
\begin{equation}
\frac{\partial \FSA{\nRE}}{\partial t} = -4\pi p_0^2\frac{\partial p_0}{\partial t}\int_{p_0}^\infty  f_0 \, \dd p.
\end{equation}
This is the hot-tail runaway rate formula implemented in \DREAM, evaluated using $p_0$ calculated according to~\eqref{eq:app:hottail pc} with collision frequencies $\nu_s = \nu_c/p^3$ and $\nu_D=(1+\Zeff)\nu_c/p^3$, using the distribution function~\eqref{eq:app:analytic hot dist}. Trapping effects are captured via the effective passing fraction which enters into the electric-field coefficient $\mathcal{D}_E$.